\documentclass[aps,pra,twocolumn,superscriptaddress,nofootinbib]{revtex4-2}

\usepackage[utf8]{inputenc}
\usepackage[T1]{fontenc}
\usepackage{amsmath,amssymb,amsfonts,bm}
\usepackage{mathrsfs}
\usepackage{physics}
\usepackage{graphicx}
\usepackage{xcolor}
\usepackage{hyperref}
\usepackage{tikz}
\begin{document}

\title{Non-equilibrium quantum field theory of the free-electron laser in Keldysh formalism}

\author{Loris Di Cairano}
\email{l.di.cairano.92@gmail.com, loris.dicairano@uni.lu}

\affiliation{Department of Physics and Materials Science, University of Luxembourg, L-1511 Luxembourg City, Luxembourg}
\date{\today}
\begin{abstract}
We develop a non-equilibrium quantum field theory of the free-electron laser based on the Preparata model, using the real-time Keldysh formalism. Starting from a microscopic Lagrangian for a relativistic electron beam coupled to a single radiation mode, we construct a Keldysh functional integral, perform the large-$N$ rescaling, and integrate out the electronic degrees of freedom. This yields an effective action for the FEL mode in which dispersion, gain, and noise are all generated by a single electronic self-energy built from the current correlations of the beam. For a stationary Gaussian beam, we obtain closed analytic expressions for the retarded and Keldysh components of the self-energy, which directly encode frequency pulling, gain reduction due to energy spread, and the noise spectrum experienced by the field. At low frequency, the theory reduces to a Landau-Ginzburg-Keldysh description of a single complex mode with a mass, growth rate, nonlinearity, and noise strength fully determined by beam current, energy spread, and detuning. In this framework, the FEL threshold appears as a continuous non-equilibrium phase transition in the laser universality class: the coherent field amplitude plays the role of an order parameter, while the amplitude of critical fluctuations is fixed by the microscopic noise kernel. The result is a minimal open quantum field theory analog of Vlasov-Maxwell FEL theory, in which gain, dispersion, and noise arise from a unified self-energy framework rather than from separate phenomenological ingredients.
\end{abstract}

\maketitle

\section{Introduction}

Free-electron lasers (FELs) have become a central tool to generate intense, coherent radiation from the infrared to the hard X-ray domain~\cite{bonifacio-collective,madey1971,colson1977,dattoli2004}. On the theoretical side, their operation is usually described by a hierarchy of models, ranging from one-dimensional Pierce-like theories~\cite{pierce1950} and pendulum equations~\cite{colson1977,bonifacio-collective} to full Vlasov--Maxwell descriptions of the electron beam and radiation field~\cite{bonifacio-gain,bonifacio-collective,kim1986}. In all these approaches, the core physical picture is clear: a relativistic electron beam moving through an undulator becomes microbunched at the radiation wavelength and transfers energy coherently to the electromagnetic field~\cite{madey1971}. The onset of this cooperative emission is identified with the FEL threshold, encoded in a dispersion relation that relates beam and undulator parameters to a complex growth rate~\cite{bonifacio-collective,kim1986}.

Despite their success, these descriptions are essentially semi-classical. The electron beam is treated as a phase-space distribution, the radiation as a classical field, and quantum effects enter only through corrections (e.g.\ recoil, energy spread) that are added on top of an otherwise classical dynamics~\cite{kim1986,saldin2000}. Noise --- which is crucial for phenomena such as self-amplified spontaneous emission (SASE)~\cite{bonifacio-gain,saldin2000,pellegrini2016}, start-up from shot noise~\cite{kim1986,saldin2000}, and statistical properties of FEL pulses~\cite{saldin2000,pellegrini2016} --- is typically introduced phenomenologically, either as white noise terms in classical equations or as stochastic initialization of the electron distribution~\cite{kim1986,saldin2000}. As a result, dispersion, gain and fluctuations are often handled by separate pieces of modeling rather than emerging from a single microscopic framework.

An alternative, fully quantum description was proposed in the pioneering work of Preparata~\cite{preparata,preparatabook,preparata2}, who formulated a quantum field theory (QFT) of the FEL in terms of a matter field for the electrons and a single bosonic mode for the radiation. In that approach, the large number of electrons $N$ is used as an expansion parameter, and the classical FEL equations appear as saddle-point equations of a quantum action. Further quantum treatments have explored collective recoil effects~\cite{bonifacio-quantum,moore1999}, quantum noise in high-gain regimes~\cite{bonifacio-quantum}, and connections to superradiance~\cite{bonifacio-gain}. However, the original Preparata formulation remained essentially deterministic: it focuses on mean-field evolution and does not systematically exploit modern tools of non-equilibrium quantum field theory, such as the Schwinger-Keldysh formalism~\cite{schwinger,keldysh,kamenev,rammer2007quantum}, to describe fluctuations, response and noise on equal footing.

In parallel, a large body of work has developed in quantum optics~\cite{carmichael,gardiner,scully1997} and open quantum systems~\cite{breuer,lindblad1976,gorini1976}, where lasers, cavity QED setups and driven--dissipative many-body systems are understood as non-equilibrium phase transitions in open quantum field theories~\cite{degiorgio,haken,diehl_review,sieberer2016keldysh,carusotto2013}. The Schwinger-Keldysh (or closed-time-path) formalism~\cite{schwinger,keldysh,chou,rammer} provides a systematic framework for treating non-equilibrium quantum dynamics, encoding causality, dissipation, and fluctuations in a unified real-time path integral~\cite{kamenev,calzetta,altland,rammer2007quantum}. Originally developed for condensed matter applications~\cite{keldysh,rammer}, it has been extended to quantum optics~\cite{carmichael,gardiner}, ultracold atoms~\cite{Berges,diehl_review}, and driven-dissipative phase transitions~\cite{sieberer2016keldysh,carusotto2013,diehl_review}. In that context, the lasing threshold is viewed as a critical point of an effective action: an ``order parameter'' field becomes unstable, a new stationary state with finite amplitude emerges, and fluctuations around the threshold can be analyzed using field-theoretic techniques from non-equilibrium statistical mechanics~\cite{hohenberg1977theory,cross_hohenberg,tauber2014critical}. This perspective has been successfully applied to optical lasers~\cite{degiorgio,haken,graham1970}, exciton-polariton condensates~\cite{carusotto2013,wouters2007}, Dicke-type models~\cite{dimer2007,nagy2010}, and a variety of driven--dissipative systems~\cite{diehl_review,sieberer2016keldysh}, but it has not been systematically brought to bear on the FEL in the language of a microscopic QFT for the electron beam and radiation.

The purpose of this work is to bridge these perspectives. We revisit the quantum FEL model of Preparata~\cite{preparata,bonifacio-quantum,preparatabook,preparata2,bonifacio-quantum} using the Schwinger-Keldysh formalism~\cite{schwinger,keldysh,kamenev} and recast it as an open quantum field theory for a single radiation mode coupled to a many-body electronic medium. Technically, we formulate a real-time Keldysh functional integral for the Preparata Lagrangian, perform the large-$N$ rescaling, and integrate out the electronic degrees of freedom to obtain an effective non-equilibrium action for the FEL mode. In this effective theory, the role played by the FEL dispersion relation in classical Vlasov--Maxwell treatments~\cite{bonifacio-collective,kim1986} is taken over by a retarded self-energy: instead of an integral over the beam energy distribution~\cite{bonifacio-collective,kim1986}, the mode dynamics is governed by an electronic self-energy expressed as a sum over single-particle levels and their occupation differences. The beam properties thus enter through well-defined correlation functions of the electronic ``current'' at the FEL wavevector, in complete analogy with the Kubo formalism for linear response in condensed matter physics~\cite{kubo1957,mahan2000}.

This construction leads naturally to a Landau-Ginzburg-Keldysh action for the FEL mode~\cite{sieberer2016keldysh,hohenberg1977theory}, in which the lasing threshold appears as a non-equilibrium phase transition. The real part of the electronic self-energy determines an effective ``mass'' parameter $r$ controlling the stability of the trivial (no-radiation) state; the imaginary part yields an effective growth/damping rate; and the Keldysh component fixes the strength of the noise~\cite{kamenev,sieberer2016keldysh}. Above threshold, a finite-amplitude stationary solution for the radiation field emerges as an order parameter, and its properties are dictated by the same microscopic self-energy that encodes gain and frequency shift. In this sense, our approach provides a minimal QFT analogue of the classical Vlasov--Maxwell theory: what is usually encoded in a dispersion integral over the beam distribution~\cite{bonifacio-collective,kim1986} is here encoded in a self-energy built from microscopic occupation numbers.

Within this framework, the FEL threshold is no longer an external condition imposed on a dispersion relation, but the point where the retarded propagator of the radiation mode acquires a zero and the effective mass $r$ changes sign --- the hallmark of a continuous phase transition in Landau theory~\cite{hohenberg1977theory}. At the same time, fluctuations and noise --- essential for SASE start-up~\cite{saldin2000,pellegrini2016} and pulse statistics~\cite{saldin2000} --- are generated consistently by the same Keldysh functional, without introducing stochastic terms by hand. This opens the way to a unified treatment of gain, saturation and fluctuations in FELs, and allows one to import concepts and tools from non-equilibrium quantum criticality~\cite{hohenberg1977theory,tauber2014critical,sieberer2016keldysh,carusotto2013} into the FEL context. In the rest of the paper we make this program explicit: we derive the Keldysh action for the Preparata model, compute the electronic self-energy for a simple beam configuration, and show how the standard notion of FEL threshold is recovered as a non-equilibrium phase transition of the effective radiation field.

\section{Keldysh formulation of the Preparata model}\label{sec:keldysh-formalism}

\subsection{The Preparata model: from many-body electron Hamiltonian to quantum field theory}

The starting point of Preparata's formulation is the $N$-electron Hamiltonian for a single-pass free-electron laser, which describes the interaction of $N$ electrons with a single radiation mode. In the ponderomotive phase representation, the Hamiltonian reads \cite{preparata,bonifacio-quantum,preparatabook,preparata2}:
\begin{equation}
    H = \sum_{j=1}^N\left[\frac{\hat{p}_j^2}{2\rho}  + i\frac{\rho^{1/2}}{N^{1/2}}\left(\hat{a}^\dagger e^{-i\theta_j} - \hat{a}e^{i\theta_j}\right)\right]-\omega_\eta \hat{a}^\dagger \hat{a},
\label{eq:preparata_hamiltonian}
\end{equation}
where $\theta_j$ is the ponderomotive phase of the $j$-th electron, $\hat{p}_j$ is its conjugate momentum, $\omega_\eta=\delta/\rho$ is the frequency with detuning $\delta$, and $\hat{a}$ ($\hat{a}^\dagger$) are the annihilation (creation) operators for the single-mode radiation field.

The dimensionless parameter $\rho$ is defined in terms of physical beam and undulator parameters as:
\begin{equation}
\rho = \frac{e I \lambda_u}{2\pi \varepsilon_0 m_e c^2 \gamma_0 \omega_0}
\label{eq:rho_def}
\end{equation}
where $I$ is the electron beam current, $\lambda_u$ is the undulator wavelength, $\gamma_0$ is the electron Lorentz factor, $\omega_0$ is the radiation frequency, $e$ is the electron charge, $m_e$ is the electron mass, $c$ is the speed of light, and $\varepsilon_0$ is the vacuum permittivity.

Preparata's key insight was to reformulate this $N$-body problem as a $(1+1)$-dimensional quantum field theory. He introduced a basis of occupation numbers $n_k$ for the electron states $|k\rangle = e^{ik\phi}/(2\pi)^{1/2}$, treating electrons as bosons in the large-$N$ limit with operators
\begin{equation}
    \hat{n}_k = \hat{c}_k^\dagger \hat{c}_k, \qquad [\hat{c}_k, \hat{c}_n^\dagger] = \delta_{kn},
\label{eq:mode_operators}
\end{equation}
and defined a complex scalar field
\begin{equation}
    \Psi(\phi,t) = \frac{1}{\sqrt{2\pi}}\sum_{k\in\mathbb{Z}} \hat{c}_k e^{ik\phi},
\label{eq:field_definition}
\end{equation}
satisfying the canonical equal-time commutation relation
\begin{equation}
    [\Psi(\phi,t), \Psi^\dagger(\phi',t)] = \delta(\phi-\phi').
\label{eq:canonical_commutation}
\end{equation}

The evolution operator $U$ associated with the Hamiltonian $H$ in Eq.~\eqref{eq:preparata_hamiltonian} generates the transition amplitude $Z:=\langle \psi_f,\alpha_f|U(t_f-t_i)|\psi_i,\alpha_i\rangle$ between coherent states of $\Psi$ and $\hat a$, which admit a path-integral representation
\[
    Z[\psi,\alpha]=\int \mathcal{D}\psi \,\mathcal{D}\alpha \,e^{i S_{T}[\psi,\alpha]}\,,
\]
with action $S_T=\int dt\,d\phi\,\mathscr{L}[\psi,\alpha]$, and Lagrangian
\[
\begin{split}
    \mathscr{L}[\psi,\alpha]=
  &\psi^*(\phi,t)\, i\partial_t \psi(\phi,t)
  + \frac{1}{2\eta}\,\psi^*(\phi,t)\,\partial_\phi^2\psi(\phi,t)
 \nonumber\\
 &- i\,\eta^{1/2}
    \Big(\alpha^*(t)e^{-i\phi} - \alpha(t)e^{i\phi}\Big)\,
    \xi^*(\phi,t)\xi(\phi,t)
  \nonumber\\
  &+ \frac{1}{2\pi}\alpha^*(t)\,i\partial_t \alpha(t)
  + \frac{\omega_\eta}{2\pi}\,\alpha^*(t)\alpha(t),
\end{split}
\]
where $\psi,\alpha$ are now coherent-state fields (c-numbers). 

The crucial step is a rescaling of the fields that isolates $N$ as an expansion parameter:
\begin{align}
    \psi(\phi,t) \to \xi(\phi,t) &:= \frac{1}{\sqrt{N}}\psi(\phi,t), \nonumber\\
    \alpha(t) \to\quad b(t) &:= \frac{1}{\sqrt{N}}\alpha(t),
\label{eq:rescaling}
\end{align}
with the normalization constraint
\begin{equation}
    \int_0^{2\pi} d\phi\, \xi^*(\phi,t)\xi(\phi,t) = 1.
\label{eq:N-constraint}
\end{equation}
This rescaling makes the action extensive in $N$, so that the large-$N$ limit plays the role of a semiclassical limit.

Inserting the rescaling~\eqref{eq:rescaling} into the original Lagrangian and using the constraint~\eqref{eq:N-constraint}, one obtains the rescaled Lagrangian density
\begin{align}
  \mathscr{L}[\xi,b]
  =
  &\xi^*(\phi,t)\, i\partial_t \xi(\phi,t)
  + \frac{1}{2\eta}\,\xi^*(\phi,t)\,\partial_\phi^2\xi(\phi,t)
 \nonumber\\
 &- i\,\eta^{1/2}
    \Big(b^*(t)e^{-i\phi} - b(t)e^{i\phi}\Big)\,
    \xi^*(\phi,t)\xi(\phi,t)
  \nonumber\\
  &+ \frac{1}{2\pi}b^*(t)\,i\partial_t b(t)
  + \frac{\omega_\eta}{2\pi}\,b^*(t)b(t),
  \label{eq:FEL-L-rescaled}
\end{align}
and the action
\begin{align}
  S_N[\xi,b]
  &= N \int_{t_0}^{t_f} dt\int_0^{2\pi} d\phi\;\mathscr{L}[\xi,b]
  \nonumber\\
  &\qquad\qquad=:S_M[\xi] + S_b[b] + S_{\mathrm{int}}[\xi,b],
  \label{eq:FEL-action-rescaled}
\end{align}
with
\begin{align}
  S_M[\xi]
  &= N \int_{t_0}^{t_f} dt\int_0^{2\pi} d\phi\,
     \bigg[
       \xi^*(\phi,t)\, i\partial_t \xi(\phi,t)
       \nonumber\\
       &\qquad\qquad\qquad\qquad+ \frac{1}{2\eta}\,\xi^*(\phi,t)\,\partial_\phi^2\xi(\phi,t)
     \bigg],
  \label{eq:SM-def-K}\\
  S_b[b]
  &= \frac{N}{2\pi}\int_{t_0}^{t_f} dt\,
     \left[
       b^*(t)\,i\partial_t b(t)
       + \omega_\eta\,b^*(t)b(t)
     \right],
  \label{eq:Sb-def-K}\\
  S_{\mathrm{int}}[\xi,b]
  &= -iN\eta^{1/2}\nonumber\\
  &\times\int_{t_0}^{t_f} dt\int_0^{2\pi} \!\!\!\!d\phi\,
     \Big(b^*(t)e^{-i\phi} - b(t)e^{i\phi}\Big)\,
     \rho(\phi,t),
  \label{eq:Sint-def-K}
\end{align}
where $\rho(\phi,t) := \xi^\dagger(\phi,t)\xi(\phi,t)$ is the electronic density. Here $S_M[\xi]$ describes the free electron dynamics in the ponderomotive potential, $S_b[b]$ the free radiation mode, and $S_{\mathrm{int}}[\xi,b]$ the electron–radiation coupling.

In the limit $N\to\infty$, the factor $N$ in front of the action implies that the functional integral is dominated by the saddle point. Varying the action yields coupled classical equations for $b_{\mathrm{cl}}(t)$ and $\xi_{\mathrm{cl}}(\phi,t)$:
\begin{align}
    \frac{db_{cl}}{dt}(t)&-i\omega_{\eta}b_{cl}(t)=\eta^{1/2}\int_{0}^{2\pi}|\xi_{cl}(\phi,t)|^{2}\,e^{-i\phi}\;d\phi,\nonumber\\
    i\frac{\partial\xi_{cl}}{\partial t}(\phi,t)&=-\frac{1}{2\eta}\frac{\partial^{2}\xi_{cl}}{\partial\phi^{2}}(\phi,t)\nonumber\\
    &\qquad+i\eta^{1/2}(b_{cl}^{*}(t)e^{-i\phi}-b_{cl}(t)\,e^{i\phi})\xi_{cl}(\phi,t).\nonumber
\end{align}
Preparata's original work focused on these mean-field equations and $O(1/\sqrt{N})$ corrections. In the present work, we instead use the full Schwinger-Keldysh formulation, which treats dynamics, response, and fluctuations on equal footing and allows us to identify the FEL threshold as a non-equilibrium phase transition.

\subsection{Position of the problem and our approach}

This framework has however two limitations. First, noise and fluctuations are not systematically included: in real FELs, shot noise in the electron beam seeds SASE and controls pulse statistics, but in the Preparata formalism, these effects must be added \textit{ad hoc}. Second, response and dissipation are not encoded in a compact way: dispersion and gain are extracted from separate evolution equations rather than from a single response kernel.

To overcome these limitations, we reformulate the Preparata model using the Schwinger-Keldysh (``in-in'') formalism, the standard real-time path-integral technique for non-equilibrium quantum systems. This approach has been successfully applied to lasers, cavity QED, driven-dissipative condensates, and open quantum systems~\cite{kamenev,sieberer2016keldysh,altland,Berges}. Its key advantage is that dynamics (equations of motion), linear response (retarded/advanced propagators), and noise (Keldysh component) all follow from a single generating functional, without introducing phenomenological stochastic terms.

In what follows, we construct the Keldysh generating functional for the Preparata model, perform the large-$N$ rescaling that isolates the collective radiation mode, and integrate out the electronic degrees of freedom to obtain an effective action for the FEL mode. The resulting self-energy $\Sigma(\omega)$ encodes dispersion, gain, and noise in a unified way.

\subsection{The Schwinger-Keldysh closed time contour}

In ordinary quantum mechanics, time evolution is generated by a unitary operator $\hat{U}(t_f, t_0)$, and expectation values are evaluated as
\begin{align}
    \langle \hat{O}(t) \rangle = \langle \psi_0 | \hat{U}^\dagger(t, t_0) \hat{O} \hat{U}(t, t_0) | \psi_0 \rangle.
\end{align}
This involves both a forward and a backward time evolution (ket and bra). The Schwinger-Keldysh formalism implements this structure by introducing a closed time contour $\mathcal{C}$ running forward from $t_0$ to $t_f$ (branch $\mathcal{C}_+$) and backward from $t_f$ to $t_0$ (branch $\mathcal{C}_-$). Any field on the contour thus has two copies:
\begin{itemize}
\item $\xi_+(\phi, t)$ for $t \in \mathcal{C}_+$ (forward branch),
\item $\xi_-(\phi, t)$ for $t \in \mathcal{C}_-$ (backward branch).
\end{itemize}
The same doubling applies to the radiation mode $b_\pm(t)$.

\subsubsection{The contour-ordered generating functional}

We define the generating functional for correlation functions by \cite{kamenev}:
\begin{widetext}
    \begin{align}
\begin{split}
Z[J_\pm, \alpha_\pm] = \text{Tr} \bigg\{ \hat{\rho}_0 \, T_\mathcal{C} \exp \bigg[ &i \int_\mathcal{C} dt \int_0^{2\pi} d\phi \, \left( J^\dagger(\phi,t) \hat{\xi}(\phi,t) + \hat{\xi}^\dagger(\phi,t) J(\phi,t) \right) + i \int_\mathcal{C} dt \left( \alpha^*(t) \hat{b}(t) + \hat{b}^\dagger(t) \alpha(t) \right) \bigg] \bigg\},
\end{split}
\end{align}
\end{widetext}
where $\hat{\rho}_0$ is the initial density matrix, $T_\mathcal{C}$ is contour ordering, and $J,\alpha$ are external sources taking different values on the two branches:
\begin{align}
J(t) = \begin{cases} J_+(\phi,t) & t \in \mathcal{C}_+, \\ J_-(\phi,t) & t \in \mathcal{C}_-, \end{cases}, \qquad \alpha(t) = \begin{cases} \alpha_+(t) & t \in \mathcal{C}_+, \\ \alpha_-(t) & t \in \mathcal{C}_-. \end{cases}
\end{align}
Correlation functions follow by functional differentiation, e.g.
\begin{align}
    \langle T_\mathcal{C} \, \hat{b}(t_1) \hat{b}^\dagger(t_2) \rangle = \frac{1}{Z[0]} \frac{\delta^2 Z}{\delta \alpha(t_1) \delta \alpha^*(t_2)} \Bigg|_{\alpha=0}.
    \label{eq:J-def-K}
\end{align}
Since the contour is closed, the trace can be written as a path integral over fields on both branches.

\subsection{Path integral representation}

Inserting coherent-state resolutions of identity along $\mathcal{C}$ \cite{kamenev,altland,calzetta}, one obtains a functional integral over complex fields $\xi_\pm(\phi, t)$ and $b_\pm(t)$:
\begin{align}
\begin{split}\label{def:keldish-action}
    Z[J_\pm, \alpha_\pm] = \int &\mathcal{D}\xi_+ \mathcal{D}\xi_- \mathcal{D}b_+ \mathcal{D}b_- \, e^{i S_\mathcal{C}[b_{\pm},\xi_{\pm}]} \\
    &\times \mathcal{P}[\xi_+(t_0), \xi_-(t_0); b_+(t_0), b_-(t_0)],
\end{split}
\end{align}
where
\begin{align}
    S_\mathcal{C}[b_{\pm},\xi_{\pm}]:=S[\xi_+, b_+] - i S[\xi_-, b_-] + i S_{\text{src}}[\xi_\pm, b_\pm; J_\pm, \alpha_\pm]
\end{align}
and $S[\xi,b]$ is the classical action from Eq.~\eqref{eq:FEL-L-rescaled}. $S_{\text{src}}$ contains the source couplings, and $\mathcal{P}$ encodes the initial density matrix via the boundary fields at $t_0$.

The difference
\begin{align}
S_K[\xi_+, \xi_-; b_+, b_-] = S[\xi_+, b_+] - S[\xi_-, b_-]
\end{align}
is the Keldysh action. At this stage the structure involves two copies of each field, with no clear physical separation between them.

\subsection{Rotation to classical/quantum (Keldysh) fields}

The standard Keldysh rotation separates average configurations from response fields:
\begin{equation}
\begin{split}\label{def:rotation-xi}
    \xi_c(\phi, t) &= \frac{\xi_+(\phi,t) + \xi_-(\phi,t)}{2}, \\ 
    \xi_q(\phi,t) &= \xi_+(\phi,t) - \xi_-(\phi,t),
\end{split}
\end{equation}
and for the radiation mode
\begin{equation}
    \begin{split}\label{def:rotation-b}
    b_c(t) &= \frac{b_+(t) + b_-(t)}{2}, \\
    b_q(t) &= b_+(t) - b_-(t).
    \end{split}
\end{equation}
The functional measure is unchanged:
\begin{align}
\mathcal{D}\xi_+ \mathcal{D}\xi_- = \mathcal{D}\xi_c \mathcal{D}\xi_q, \qquad \mathcal{D}b_+ \mathcal{D}b_- = \mathcal{D}b_c \mathcal{D}b_q.
\end{align}

In this basis, $b_c$ and $\xi_c$ represent the physical (average) fields, while $b_q,\xi_q$ encode how observables change under perturbations. More concretely, $b_c(t)$ is the coherent amplitude of the radiation mode, and $b_q(t)$ acts as a Lagrange multiplier enforcing the equations of motion for $b_c(t)$ when one varies the action with respect to $b_q^\dagger$.

\subsection{Structure of the Keldysh action in $(c,q)$ basis}

Substituting Eqs.~\eqref{def:rotation-xi} and \eqref{def:rotation-b} into Eq.~\eqref{def:keldish-action} and expanding to leading order in the quantum fields (sufficient for Gaussian fluctuations and linear response), the action splits into matter, radiation, and interaction pieces (see Appendix~\ref{app:keldysh-rotation} for details).

\subsubsection{Free matter part}
\begin{align}
\begin{split}
    S_{M,K}[\xi_c, \xi_q] = N \int_{t_0}^{t_f} dt &\int_0^{2\pi} d\phi\\
    \times\bigg[ \xi_q^\dagger(\phi,t) \mathcal{L}&\,\xi_c(\phi,t) + \xi_c^\dagger(\phi,t) \mathcal{L}\,\xi_q(\phi,t) \bigg],
\end{split}
\end{align}
with
\[
     \mathcal{L}:=i\partial_t + \frac{1}{2\eta} \partial_\phi^2.
\]
The action is linear in $\xi_q$, which enforces the matter equation of motion
\begin{align}
    \left( i\partial_t + \frac{1}{2\eta} \partial_\phi^2 \right) \xi_c(\phi,t) = \text{(interaction terms)}.
\end{align}

\subsubsection{Free radiation part}
\begin{align}
    S_{b,K}[b_c, b_q] = \frac{N}{2\pi} \int_{t_0}^{t_f} dt \left[ b_q^\dagger(t) \mathcal{D}\,b_c(t) + b_c^\dagger(t) \mathcal{D}\, b_q(t) \right],
\label{eq:free_radiation}
\end{align}
with
\[
    \mathcal{D}:=i\partial_t + \omega_\eta.
\]
Again, only mixed $b_q^\dagger b_c$ and $b_c^\dagger b_q$ terms appear.

\subsubsection{Interaction part}

The interaction couples the radiation to the density $\rho(\phi,t) = \xi^\dagger(\phi,t) \xi(\phi,t)$. On each branch,
\begin{align}
\rho_\pm(\phi,t) = \xi_\pm^\dagger(\phi,t) \xi_\pm(\phi,t),
\end{align}
and we define
\[
     \rho_c = \frac{\rho_+ + \rho_-}{2}, \quad \rho_q = \rho_+ - \rho_-.
\]
To leading order
\begin{align}
\rho_c \approx \xi_c^\dagger \xi_c, \qquad \rho_q \approx \xi_q^\dagger \xi_c + \xi_c^\dagger \xi_q.
\end{align}
The interaction action takes the form
\begin{align}
    S_{\text{int},K}[\xi_c, \xi_q; b_c, b_q] = -iN\eta^{1/2} \int dt& \int d\phi \\
    \times\bigg[ ( b_c^\dagger e^{-i\phi} - b_c e^{i\phi} ) \rho_q + ( &b_q^\dagger e^{-i\phi} - b_q e^{i\phi} ) \rho_c \bigg].\nonumber
\end{align}
Thus the classical radiation field $b_c$ couples to the quantum density $\rho_q$, while $b_q$ couples to $\rho_c$. This cross-coupling between classical and quantum components is the origin of dissipation and noise in the effective theory.

\subsection{Electronic current at the FEL wavevector}

It is convenient to introduce the Fourier component of the density at wavevector $k=1$:
\begin{equation}\label{eq:current_definition}
\begin{split}
    J(t) = \int_0^{2\pi} d\phi \, &e^{-i\phi} \rho(\phi,t) \\
    &= \int_0^{2\pi} d\phi \, e^{-i\phi} \xi^\dagger(\phi,t) \xi(\phi,t).
\end{split}
\end{equation}
$J(t)$ is the microscopic bunching parameter at the radiation wavelength, the quantum analogue of the classical FEL bunching $b_1 = \langle e^{i\phi} \rangle$.

In the $(c,q)$ basis:
\begin{align}
    J_c(t) = \int_0^{2\pi} d\phi \, e^{-i\phi} \rho_c(\phi,t), \\
    J_q(t) = \int_0^{2\pi} d\phi \, e^{-i\phi} \rho_q(\phi,t).
\end{align}
The interaction action becomes
\begin{align}\label{def:S-int-K}
    S_{\text{int},K}[\xi_c, \xi_q&; b_c, b_q] = -iN\eta^{1/2} \\
    &\times\int dt\; [ b_c^\dagger(t) J_q(t) - b_c(t) J_q^\dagger(t) \nonumber\\
    &\quad\quad\qquad+ b_q^\dagger(t) J_c(t) - b_q(t) J_c^\dagger(t) ].\nonumber
\end{align}
The quantum current $J_q$ is sourced by $b_c$, while $J_c$ is sourced by $b_q$.

\subsection{Integrating out the electrons: the influence functional}

Integrating out the electronic fields produces an influence functional for the radiation:
\begin{align*}
    e^{i S_{\text{IF}}[b_c, b_q]} = \int \mathcal{D}\xi_c \mathcal{D}\xi_q \, e^{ i S_{M,K}[\xi_c, \xi_q] + i S_{\text{int},K}[\xi_c, \xi_q; b_c, b_q] }.\label{def:action-IF}
\end{align*}
The effective action for the FEL mode is then
\begin{align}
    S_{\text{eff}}[b_c, b_q] = S_{b,K}[b_c, b_q] + S_{\text{IF}}[b_c, b_q].
\end{align}
In general $S_{\mathrm{IF}}$ is non-local and non-linear in $(b_c,b_q)$. Near threshold, where the field amplitude is small, we expand $S_{\mathrm{IF}}$ in powers of $b_c,b_q$. A cumulant expansion of the current yields, to quadratic order in $b$,
\begin{equation*}
  S_{\mathrm{IF}}^{(2)}[b_c,b_q]
  = -\frac{1}{2}
    \left\langle
      \left(
        S_{\mathrm{int},K}[\xi_c,\xi_q;b_c,b_q]
      \right)^2
    \right\rangle_{0,c}
  + \mathcal{O}(b^3),
  \label{eq:SIF2-cumulant-K}
\end{equation*}
where $\langle\cdots\rangle_{0,c}$ denotes the connected average with respect to the free matter action $S_{M,K}$. If the initial beam is unbunched, $\langle J(t)\rangle_0=0$, the first-order term vanishes.

Using Eq.~\eqref{def:S-int-K} and Wick's theorem (see Appendix~\ref{app:wick}), the quadratic influence action can be written in the standard $2 \times 2$ Keldysh matrix form:
\begin{widetext}
    \begin{align}
    S_{\text{IF}}^{(2)}[b_c, b_q] = -\int dt \, dt' \begin{pmatrix} b_q^\dagger(t) & b_c^\dagger(t) \end{pmatrix} \begin{pmatrix} 0 & \Sigma^A(t,t') \\ \Sigma^R(t,t') & \Sigma^K(t,t') \end{pmatrix} \begin{pmatrix} b_c(t') \\ b_q(t') \end{pmatrix},
    \label{eq:influence_functional}
\end{align}
\end{widetext}
where the retarded, advanced, and Keldysh self-energies are
\begin{align}
    \Sigma^R(t,t') &= -iN\eta \, \theta(t-t') \left\langle \left[ J(t), J^\dagger(t') \right] \right\rangle_0, \\
    \Sigma^A(t,t') &= +iN\eta \, \theta(t'-t) \left\langle \left[ J(t), J^\dagger(t') \right] \right\rangle_0, \\
    \Sigma^K(t,t') &= -iN\eta \left\langle \left\{ J(t), J^\dagger(t') \right\} \right\rangle_0.
\end{align}
$\Sigma^R$ encodes the retarded response of the current to the field (dispersion and gain), $\Sigma^A$ its advanced counterpart, and $\Sigma^K$ the current fluctuation spectrum (noise). The three components are not independent: they satisfy a fluctuation–dissipation structure dictated by the Keldysh formalism.

\subsection{Effective Keldysh action for the FEL mode}

Combining Eq.~\eqref{eq:free_radiation} with Eq.~\eqref{eq:influence_functional}, the quadratic effective action reads:
\begin{widetext}
    \begin{align}\label{def:S-eff-2}
    S_{\text{eff}}^{(2)}[b_c, b_q] = \int dt \, dt' \begin{pmatrix} b_q^\dagger(t) & b_c^\dagger(t) \end{pmatrix} \begin{pmatrix} 0 & [G_0^A]^{-1}(t,t') - \Sigma^A(t,t') \\ [G_0^R]^{-1}(t,t') - \Sigma^R(t,t') & -\Sigma^K(t,t') \end{pmatrix} \begin{pmatrix} b_c(t') \\ b_q(t') \end{pmatrix},
\end{align}
\end{widetext}
with free inverse propagators
\begin{align}
    [G_0^R]^{-1}(t,t') &= \frac{N}{2\pi} (i\partial_t + \omega_\eta) \delta(t-t'), \\
    [G_0^A]^{-1}(t,t') &= \frac{N}{2\pi} (-i\partial_t + \omega_\eta) \delta(t-t').
\end{align}
The dressed retarded/advanced inverse propagators are
\begin{align}
    [G^R]^{-1}(t,t') = [G_0^R]^{-1}(t,t') - \Sigma^R(t,t'), \\
    [G^A]^{-1}(t,t') = [G_0^A]^{-1}(t,t') - \Sigma^A(t,t').
\end{align}
In frequency space, the FEL dispersion relation is the condition that the retarded propagator has a pole,
\begin{align}
     \Gamma^R(\omega) := [G_0^R(\omega)]^{-1} - \Sigma^R(\omega) = 0,
\end{align}
which plays the role of the classical dispersion relation $D(\omega,k)=0$, but now embedded in a Keldysh matrix where fluctuations and noise are automatically incorporated through $\Sigma^K$.

In summary, starting from the Preparata Lagrangian, we constructed a real-time Keldysh path integral, performed the classical/quantum rotation, and integrated out the electrons to obtain an effective action for the single FEL mode in terms of a self-energy $\Sigma^{R,A,K}(t,t')$ built from current correlators. In the next section we compute these self-energies explicitly for a stationary beam and show how the FEL threshold emerges as the condition $\Gamma^R(\omega_{\text{res}})=0$.

\section{Electronic self-energy for a stationary beam}

We now compute the self-energies $\Sigma^{R,A,K}$, starting from the definitions
\begin{align}
\Sigma^R(t,t') &= -iN\eta \, \theta(t-t') \left\langle \left[ J(t), J^\dagger(t') \right] \right\rangle_0, \label{eq:sigma_R_time} \\
\Sigma^A(t,t') &= +iN\eta \, \theta(t'-t) \left\langle \left[ J(t), J^\dagger(t') \right] \right\rangle_0, \label{eq:sigma_A_time} \\
\Sigma^K(t,t') &= -iN\eta \left\langle \left\{ J(t), J^\dagger(t') \right\} \right\rangle_0, \label{eq:sigma_K_time}
\end{align}
for a stationary beam with diagonal initial state in the ponderomotive basis. We assume an initially unbunched beam, $\langle J(t)\rangle_0=0$, so that the first cumulant is absent.

\subsection{Mode expansion of the matter field}

The free matter Hamiltonian reads
\begin{align}
\hat{H}_M = \int_0^{2\pi} d\phi \, \hat{\xi}^\dagger(\phi) \left( -\frac{1}{2\eta} \frac{\partial^2}{\partial\phi^2} \right) \hat{\xi}(\phi),
\label{eq:matter_hamiltonian}
\end{align}
describing a quantum particle on a circle with ``mass'' $\eta$. In Fourier modes
\begin{equation}
    \begin{split}
        \hat{\xi}(\phi) &= \frac{1}{\sqrt{2\pi}} \sum_{m \in \mathbb{Z}} \hat{c}_m e^{im\phi}, \\
        \hat{\xi}^\dagger(\phi) &= \frac{1}{\sqrt{2\pi}} \sum_{m \in \mathbb{Z}} \hat{c}_m^\dagger e^{-im\phi},
\label{eq:mode_expansion}
    \end{split}
\end{equation}
with fermionic anticommutation relations
\begin{align}
\{\hat{c}_m, \hat{c}_n^\dagger\} = \delta_{mn}, \qquad \{\hat{c}_m, \hat{c}_n\} = 0.
\label{eq:anticommutation}
\end{align}
The normalization condition~(2) becomes
\begin{align}
\int_0^{2\pi} d\phi \, \hat{\xi}^\dagger(\phi,t) \hat{\xi}(\phi,t) = \sum_m \hat{c}_m^\dagger \hat{c}_m = \hat{N}_{\text{tot}},
\label{eq:normalization_modes}
\end{align}
with $\langle \hat{N}_{\text{tot}} \rangle = 1$ in the rescaled theory (the physical electron number is carried by $N$).

Inserting Eq.~\eqref{eq:mode_expansion} into Eq.~\eqref{eq:matter_hamiltonian} yields
\begin{align}
\hat{H}_M = \sum_{m \in \mathbb{Z}} \varepsilon_m \hat{c}_m^\dagger \hat{c}_m, \qquad \varepsilon_m = \frac{m^2}{2\eta}.
\label{eq:matter_spectrum}
\end{align}
The mode operators evolve as
\begin{align}
    \hat{c}_m(t) &= \hat{c}_m e^{-i\varepsilon_m t}, \qquad
    \hat{c}_m^\dagger(t) = \hat{c}_m^\dagger e^{+i\varepsilon_m t}.
\label{eq:mode_time_evolution}
\end{align}

\subsection{Electronic current at the FEL wavevector}

The density operator is
\begin{align}
\hat{\rho}(\phi) = \hat{\xi}^\dagger(\phi) \hat{\xi}(\phi) = \frac{1}{2\pi} \sum_{m,n} \hat{c}_m^\dagger \hat{c}_n e^{i(n-m)\phi}.
\label{eq:density_operator}
\end{align}
The current at $k=1$ (see Eq.~\eqref{eq:current_definition}) becomes
\begin{align}
    \hat{J} = \int_0^{2\pi}& d\phi \, e^{-i\phi} \hat{\rho}(\phi) \nonumber\\
    &= \sum_{m,n} \hat{c}_m^\dagger \hat{c}_n \frac{1}{2\pi} \int_0^{2\pi} d\phi \, e^{i(n-m-1)\phi} \nonumber\\
    &= \sum_{m \in \mathbb{Z}} \hat{c}_m^\dagger \hat{c}_{m+1}.
\label{eq:current_operator_modes}
\end{align}
$\hat{J}$ transfers population between neighboring modes $m$ and $m+1$, the quantum counterpart of the classical bunching mechanism.

Using Eq.~\eqref{eq:mode_time_evolution}, the current evolves as
\begin{align}\label{eq:current_time_evolution}
    \hat{J}(t) &= \sum_m \hat{c}_m^\dagger(t) \hat{c}_{m+1}(t) \\
    &= \sum_m \hat{c}_m^\dagger \hat{c}_{m+1} e^{i(\varepsilon_m - \varepsilon_{m+1})t} = \sum_m \hat{c}_m^\dagger \hat{c}_{m+1} e^{-i\Omega_m t},\nonumber
\end{align}
with transition frequencies
\begin{align}
\Omega_m := \varepsilon_{m+1} - \varepsilon_m = \frac{2m+1}{2\eta}.
\label{eq:transition_frequency}
\end{align}

\subsection{Retarded self-energy: response and gain}

We assume a stationary, diagonal initial state with mode occupations
\begin{align}
n_m := \langle \hat{c}_m^\dagger \hat{c}_m \rangle_0,
\label{eq:occupation_numbers}
\end{align}
and no coherences: $\langle \hat{c}_m^\dagger \hat{c}_n \rangle_0 = n_m \delta_{mn}$.
The retarded current–current correlator is
\begin{align}
\chi^R_{JJ}(t) = -i \theta(t) \langle [\hat{J}(t), \hat{J}^\dagger(0)] \rangle_0.
\label{eq:retarded_correlator}
\end{align}
Using Eq.~\eqref{eq:current_time_evolution},
\begin{align}
[\hat{J}(t), \hat{J}^\dagger(0)] = \sum_{m,n} e^{-i\Omega_m t} [\hat{c}_m^\dagger \hat{c}_{m+1}, \hat{c}_{n+1}^\dagger \hat{c}_n].
\label{eq:current_commutator}
\end{align}
The commutator evaluates to
\begin{align}
\begin{split}
[\hat{c}_m^\dagger \hat{c}_{m+1}, \hat{c}_{n+1}^\dagger \hat{c}_n] &= \delta_{mn} (\hat{n}_m - \hat{n}_{m+1}),
\end{split}
\label{eq:commutator_calculation}
\end{align}
so that
\begin{align}
[\hat{J}(t), \hat{J}^\dagger(0)] = \sum_m e^{-i\Omega_m t} (\hat{n}_m - \hat{n}_{m+1}),
\label{eq:commutator_result}
\end{align}
and
\begin{align}
\langle [\hat{J}(t), \hat{J}^\dagger(0)] \rangle_0 = \sum_m (n_m - n_{m+1}) e^{-i\Omega_m t}.
\label{eq:commutator_average}
\end{align}
Using Eq.~\eqref{eq:sigma_R_time},
\begin{align}
    \Sigma^R(t) \equiv \Sigma^R(t,0) &= -iN\eta \, \theta(t) \sum_m (n_m - n_{m+1}) e^{-i\Omega_m t}.
\label{eq:sigma_R_time_explicit}
\end{align}

\subsection{Fourier transform and spectral decomposition}

Fourier transforming
\begin{align}
\Sigma^R(\omega) = \int_{-\infty}^{+\infty} dt \, e^{i\omega t} \Sigma^R(t),
\label{eq:fourier_convention}
\end{align}
and using
\begin{align}
\int_0^\infty dt \, e^{i(\omega - \Omega)t} = \pi\delta(\omega - \Omega) + i\mathcal{P}\frac{1}{\omega - \Omega},
\label{eq:fourier_identity}
\end{align}
one finds
\begin{align}
\begin{split}
    \Sigma^R(\omega) &= N\eta \sum_m (n_m - n_{m+1}) \left[ \mathcal{P}\frac{1}{\omega - \Omega_m} - i\pi\delta(\omega - \Omega_m) \right].
\end{split}
\label{eq:sigma_R_frequency}
\end{align}
Thus
\begin{align}
\text{Re}\, \Sigma^R(\omega) &= N\eta \sum_m (n_m - n_{m+1}) \mathcal{P}\frac{1}{\omega - \Omega_m}, \label{eq:sigma_R_real} \\
\text{Im}\, \Sigma^R(\omega) &= -N\eta\pi \sum_m (n_m - n_{m+1}) \delta(\omega - \Omega_m). \label{eq:sigma_R_imag}
\end{align}
Only differences $(n_m-n_{m+1})$ enter, the discrete analogue of a slope $\partial f/\partial p$ in classical FEL theory: positive slope yields gain, negative slope damping.

\subsection{Keldysh self-energy: noise spectrum}

The Keldysh component follows from the anticommutator in Eq.~\eqref{eq:sigma_K_time}. A straightforward calculation gives
\begin{align}
\langle \hat{J}(t) \hat{J}^\dagger(0) \rangle_0 &= \sum_m n_m(1 - n_{m+1}) e^{-i\Omega_m t}, \label{eq:JJ_correlator} \\
\langle \hat{J}^\dagger(0) \hat{J}(t) \rangle_0 &= \sum_m n_{m+1}(1 - n_m) e^{-i\Omega_m t}, \label{eq:JJ_correlator_reverse}
\end{align}
so
\begin{align}
\langle \{\hat{J}(t), \hat{J}^\dagger(0)\} \rangle_0 &= \sum_m S_m e^{-i\Omega_m t}, \nonumber\\
S_m &:= n_m + n_{m+1} - 2n_m n_{m+1}.
\label{eq:anticommutator_result}
\end{align}
In the dilute regime $n_m\ll 1$, $S_m \approx n_m+n_{m+1}\approx 2n(m)$. Fourier transforming yields
\begin{align}
    \Sigma^K(\omega) &= -iN\eta \sum_m S_m \cdot 2\pi\delta(\omega - \Omega_m) \\
    &\approx -i4\pi N\eta \sum_m n_m \delta(\omega - \Omega_m).
\label{eq:sigma_K_frequency}
\end{align}
$\Sigma^K(\omega)$ measures the current fluctuation spectrum and thus the strength of spontaneous emission noise driving the radiation mode.

\subsection{Threshold condition in terms of mode occupations}

The full retarded inverse propagator of the FEL mode is
\begin{align}
    \Gamma^R(\omega) &= [G_0^R(\omega)]^{-1} - \Sigma^R(\omega) 
    \nonumber\\
    &= \frac{N}{2\pi}(\omega - \omega_\eta + i0^+) - \Sigma^R(\omega).
\label{eq:full_propagator}
\end{align}
The FEL threshold corresponds to a zero of $\Gamma^R(\omega)$, i.e. to a marginally stable mode. Near threshold we look for a resonant frequency $\omega_{\text{res}}$ such that
\begin{align}
\text{Re}\, \Gamma^R(\omega_{\text{res}}) = 0.
\label{eq:threshold_condition}
\end{align}
Using Eq.~\eqref{eq:sigma_R_real} this gives
\begin{align}
\frac{N}{2\pi}(\omega_{\text{res}} - \omega_\eta) = N\eta \sum_m (n_m - n_{m+1}) \mathcal{P}\frac{1}{\omega_{\text{res}} - \Omega_m},
\label{eq:threshold_mode_occupations}
\end{align}
or
\begin{align}
\omega_{\text{res}} - \omega_\eta = 2\pi\eta \sum_m (n_m - n_{m+1}) \mathcal{P}\frac{1}{\omega_{\text{res}} - \Omega_m}.
\label{eq:threshold_implicit}
\end{align}
This is the self-energy analogue of the FEL dispersion relation. The imaginary part
\begin{align}
    \text{Im}\, \Gamma^R(\omega_{\text{res}}) &= -\text{Im}\, \Sigma^R(\omega_{\text{res}}) \nonumber\\
    &= N\eta\pi \sum_m (n_m - n_{m+1}) \delta(\omega_{\text{res}} - \Omega_m)
\label{eq:gain_formula}
\end{align}
gives the net gain/damping rate and explicitly shows that only resonant transitions and their population differences contribute.

In the next section, we evaluate these expressions for a Gaussian beam and obtain closed-form results for $\Sigma^{R,K}(\omega)$ and the threshold.

\section{Gaussian beam model: closed-form results}\label{sec:gaussian-beam}

We now specialize to a simple model of a stationary beam with finite energy spread, described by Gaussian mode occupations. This allows us to obtain explicit analytic expressions for the self-energy and make contact with standard FEL results.

\subsection{Gaussian occupations and continuum limit in mode space}

We take a diagonal state with
\begin{align}
    n_m = \frac{1}{\mathcal{N}} \exp\left[ -\frac{(m - m_0)^2}{2\sigma_m^2} \right], \nonumber\\
    \mathcal{N} = \sum_{m \in \mathbb{Z}} \exp\left[ -\frac{(m - m_0)^2}{2\sigma_m^2} \right],
\label{eq:gaussian_occupations}
\end{align}
centered around $m_0\gg 1$ with width $\sigma_m\gg 1$, so that sums over $m$ can be approximated by integrals. The energies and transition frequencies are
\begin{align}
\varepsilon_m = \frac{m^2}{2\eta}, \qquad \Omega_m = \frac{2m+1}{2\eta},
\label{eq:energies_frequencies}
\end{align}
and we define the inverse relation
\begin{align}
m(\Omega) \approx \eta\Omega - \frac{1}{2}.
\label{eq:inverse_relation}
\end{align}
For a smooth profile
\begin{align}
n_{m+1} - n_m \approx \frac{\partial n}{\partial m}\bigg|_m, \qquad n_m - n_{m+1} \approx -\frac{\partial n}{\partial m}\bigg|_m.
\label{eq:finite_difference}
\end{align}
For the Gaussian
\begin{align}
    \frac{\partial n}{\partial m} &\approx -\frac{(m - m_0)}{\sigma_m^2} n(m)
    \\
    &\Rightarrow n_m - n_{m+1} \approx \frac{(m - m_0)}{\sigma_m^2} n(m).
\label{eq:gaussian_derivative}
\end{align}

\subsection{Imaginary part of the self-energy: gain with energy spread}

Starting from Eq.~\eqref{eq:sigma_R_imag},
\begin{align}
\text{Im}\, \Sigma^R(\omega) = -N\eta\pi \sum_m (n_m - n_{m+1}) \delta(\omega - \Omega_m),
\label{eq:imag_sigma_start}
\end{align}
we take the continuum limit $\sum_m\to\int dm$ and use
\begin{align}
\delta(\omega - \Omega_m) = \eta \, \delta(m - m_{\text{res}}), \qquad m_{\text{res}} \approx \eta\omega - \frac{1}{2}.
\label{eq:delta_function_identity}
\end{align}
This yields
\begin{align}
    \text{Im}\, \Sigma^R(\omega) &\approx -N\eta^2\pi \, (n_m - n_{m+1})\big|_{m = m_{\text{res}}}.
\label{eq:imag_sigma_integral}
\end{align}
Using Eq.~\eqref{eq:gaussian_derivative} and defining
\begin{align}
\Omega_0 := \Omega_{m_0} = \frac{2m_0 + 1}{2\eta}, \qquad \sigma_\Omega := \frac{\sigma_m}{\eta}, \qquad y := \frac{\omega - \Omega_0}{\sqrt{2}\sigma_\Omega},
\label{eq:dimensionless_detuning}
\end{align}
we have
\begin{align}
n(m_{\text{res}}) = \frac{1}{\sqrt{2\pi}\sigma_m} e^{-y^2},
\label{eq:gaussian_at_resonance}
\end{align}
and obtain
\begin{align}
    \text{Im}\, \Sigma^R(\omega) \approx -N\sqrt{\pi} \, \frac{y}{\sigma_\Omega^2} e^{-y^2}.
\label{eq:imag_sigma_final}
\end{align}
The gain is proportional to $y e^{-y^2}$ and suppressed by $1/\sigma_\Omega^2$, capturing both detuning and energy spread effects.

\subsection{Real part of the self-energy: Hilbert transform and Dawson function}

For the real part,
\begin{widetext}
    \begin{align}
\text{Re}\, \Sigma^R(\omega) = N\eta \sum_m (n_m - n_{m+1}) \mathcal{P}\frac{1}{\omega - \Omega_m} \approx N\eta \int dm \, \frac{(m - m_0)}{\sigma_m^2} n(m) \mathcal{P}\frac{1}{\delta\omega - (m - m_0)/\eta},
\label{eq:real_sigma_start}
\end{align}
\end{widetext}
with $\delta\omega=\omega-\Omega_0$. Using $x=\delta m/(\sqrt{2}\sigma_m)$ and the standard Hilbert transform involving the Dawson function $F(y)$, one finds
\begin{align}
\text{Re}\, \Sigma^R(\omega) \approx \frac{N}{\sigma_\Omega^2} \left[ 2y F(y) - 1 \right],
\label{eq:real_sigma_final}
\end{align}
with
\begin{align}
F(y) = e^{-y^2} \int_0^y dt \, e^{t^2}.
\label{eq:dawson_function}
\end{align}
Altogether,
\begin{align}
    \Sigma^R(\omega) &\approx \frac{N}{\sigma_\Omega^2} \big[ 2y F(y) - 1 \big] - iN\sqrt{\pi} \, \frac{y}{\sigma_\Omega^2} e^{-y^2},
\label{eq:sigma_R_complete}
\end{align}
with $y$ given above. Frequency pulling is encoded in $2yF(y)-1$, and gain in $y e^{-y^2}$.

\subsection{Dressed mode, threshold condition and gain}

The retarded inverse propagator becomes
\begin{align}
    \Gamma^R(\omega) \approx N\left[ \frac{\omega - \omega_\eta}{2\pi} - \frac{1}{\sigma_\Omega^2}\left(2yF(y) - 1\right) + i\frac{\sqrt{\pi}}{\sigma_\Omega^2} y e^{-y^2} \right].
\label{eq:gamma_R_explicit}
\end{align}
The real part gives the dispersion relation
\begin{align}
\frac{\omega_{\text{res}} - \omega_\eta}{2\pi} = \frac{1}{\sigma_\Omega^2}\left[ 2y_{\text{res}} F(y_{\text{res}}) - 1 \right],
\label{eq:dispersion_relation}
\end{align}
while the imaginary part evaluated at $\omega_{\text{res}}$ gives the gain
\begin{align}
\text{Im}\, \Gamma^R(\omega_{\text{res}}) = N\frac{\sqrt{\pi}}{\sigma_\Omega^2} y_{\text{res}} e^{-y_{\text{res}}^2}.
\label{eq:gain_formula_gaussian}
\end{align}
As in standard FEL theory, gain is maximal on the falling side of the distribution and suppressed by energy spread.

\subsection{Noise spectrum and effective occupation}

Applying the same continuum approximation to Eq.~\eqref{eq:sigma_K_frequency}, we obtain
\begin{align}
\Sigma^K(\omega) \approx -i \frac{4\sqrt{2\pi} N\eta}{\sigma_\Omega} e^{-y^2}.
\label{eq:sigma_K_final}
\end{align}
Comparing with Eq.~\eqref{eq:imag_sigma_final}, the Keldysh relation
\begin{align}
\Sigma^K(\omega) = 2i F(\omega) \, \text{Im}\, \Sigma^R(\omega)
\label{eq:keldysh_relation}
\end{align}
defines an effective occupation $F(\omega)$,
\begin{align}
F(\omega) = \frac{\Sigma^K(\omega)}{2i\, \text{Im}\, \Sigma^R(\omega)} \approx \frac{2\sqrt{2}\eta\sigma_\Omega}{y},
\label{eq:effective_occupation}
\end{align}
which has a strong frequency dependence and cannot be reduced to a simple thermal form. The beam acts as a highly non-equilibrium bath whose effective ``temperature'' diverges near the center of the distribution and vanishes far away.

Summarizing, for the Gaussian beam we have
\begin{align}
    \text{Re}\, \Sigma^R(\omega) &\approx \frac{N}{\sigma_\Omega^2} \left[ 2y F(y) - 1 \right], \label{eq:summary_real} \\
    \text{Im}\, \Sigma^R(\omega) &\approx -N\sqrt{\pi} \, \frac{y}{\sigma_\Omega^2} e^{-y^2}, \label{eq:summary_imag} \\
    \Sigma^K(\omega) &\approx -i \frac{4\sqrt{2\pi} N\eta}{\sigma_\Omega} e^{-y^2}, \label{eq:summary_keldysh}
\end{align}
with $y = (\omega - \Omega_0)/(\sqrt{2}\sigma_\Omega)$ and $F(y)$ the Dawson integral.

\section{Effective Landau-Ginzburg-Keldysh theory and universality}\label{sec:LGK-universality}

We now use these results to derive a low-frequency, small-amplitude Landau-Ginzburg-Keldysh (LGK) theory for the FEL mode and identify the universality class of the threshold.

\subsection{Low-frequency expansion of the dressed propagator}

We Fourier transform fields and self-energies:
\begin{align}
    b_{c,q}(t) &= \int \frac{d\omega}{2\pi} e^{-i\omega t} b_{c,q}(\omega), \nonumber\\
    \Sigma^{R,A,K}(t,t') &= \int \frac{d\omega}{2\pi} e^{-i\omega(t-t')} \Sigma^{R,A,K}(\omega).
\label{eq:fourier_transform_fields}
\end{align}
Then Eq.~\eqref{def:S-eff-2} becomes
\begin{widetext}
\begin{align}
    S_{\text{eff}}^{(2)}[b_c, b_q] = \int \frac{d\omega}{2\pi} \begin{pmatrix} b_q^\dagger(\omega) & b_c^\dagger(\omega) \end{pmatrix} \begin{pmatrix} 0 & [G_0^A(\omega)]^{-1} - \Sigma^A(\omega) \\ [G_0^R(\omega)]^{-1} - \Sigma^R(\omega) & -\Sigma^K(\omega) \end{pmatrix} \begin{pmatrix} b_c(\omega) \\ b_q(\omega) \end{pmatrix},
\label{eq:effective_action_frequency}
\end{align}
\end{widetext}
with
\begin{align}
    [G_0^R(\omega)]^{-1} &= \frac{N}{2\pi}(\omega - \omega_\eta + i0^+), \\
    [G_0^A(\omega)]^{-1} &= \frac{N}{2\pi}(\omega - \omega_\eta - i0^+).
\label{eq:free_propagators_frequency}
\end{align}
We define
\begin{align}
\Gamma^R(\omega) = [G_0^R(\omega)]^{-1} - \Sigma^R(\omega).
\label{eq:full_retarded_propagator}
\end{align}
In a rotating frame around the relevant frequency we expand at low $\omega$:
\begin{align}
\Gamma^R(\omega) \approx Z^{-1} \left[ \omega - \Delta\omega + i\kappa \right],
\label{eq:low_frequency_expansion}
\end{align}
with
\begin{align}
Z^{-1} &= \frac{\partial \Gamma^R(\omega)}{\partial \omega}\bigg|_{\omega=0}, \label{eq:Z_factor} \\
\Delta\omega &= \frac{\text{Re}\, \Gamma^R(0)}{Z^{-1}}, \label{eq:frequency_shift} \\
\kappa &= -\frac{\text{Im}\, \Gamma^R(0)}{Z^{-1}}. \label{eq:growth_rate}
\end{align}
Here $Z$ rescale time derivatives, $\Delta\omega$ is a frequency shift (frequency pulling), and $\kappa$ is the linear growth/damping rate. From $\Sigma^K(\omega)$ we define a noise strength
\begin{align}
D = \frac{1}{2} \Sigma^K(\omega \approx 0),
\label{eq:noise_strength}
\end{align}
determined by beam parameters.

Transforming back to time and keeping only the leading derivative (see Appendix~\ref{app:low_frequency}), the quadratic action reduces to
\begin{align}
    S_{\text{eff}}^{(2)}[b_c, b_q] &\approx \int dt \, \bigg\{ b_q^\dagger(t) [ Z i\partial_t - r + i\kappa ] b_c(t) \label{eq:effective_action_time}\\
    &+ b_c^\dagger(t) [ Z i\partial_t - r - i\kappa ] b_q(t) + iD |b_q(t)|^2 \bigg\},
\nonumber
\end{align}
where
\begin{align}
r = Z \Delta\omega
\label{eq:mass_parameter}
\end{align}
is an effective ``mass'' that changes sign at threshold and parametrizes the distance from it.

\subsection{Nonlinear saturation and Landau-Ginzburg-Keldysh action}

Nonlinear saturation arises from higher-order terms in $S_{\text{IF}}$, generating cubic vertices of the form
\begin{align}
S_{\text{eff}}^{(3)}[b_c, b_q] \sim -\int dt \, \left[ \lambda |b_c(t)|^2 b_q^\dagger(t) + \lambda^* |b_c(t)|^2 b_q(t) \right],
\label{eq:cubic_interaction}
\end{align}
with complex $\lambda$ determined by three-point current correlators. Collecting quadratic and cubic contributions we obtain the LGK action
\begin{widetext}
\begin{align}
    S_{\text{LGK}}[b_c, b_q] = \int dt \, \left\{ b_q^\dagger \left[ Z i\partial_t - r + i\kappa \right] b_c + b_c^\dagger \left[ Z i\partial_t - r - i\kappa \right] b_q + iD |b_q|^2 - \lambda |b_c|^2 b_q^\dagger - \lambda^* |b_c|^2 b_q + \cdots \right\},
\label{eq:LGK_action}
\end{align}
\end{widetext}
where dots denote higher-order and higher-derivative terms that are irrelevant for the low-frequency dynamics near threshold.

\subsection{Effective Langevin equation and phase transition}

Varying $S_{\text{LGK}}$ with respect to $b_q^\dagger$ \cite{Maghrebi} gives the equation of motion
\begin{align}
    \frac{\delta S_{\text{LGK}}}{\delta b_q^\dagger(t)} = 0 \,,
\end{align}
namely
\begin{align}\label{eq:equation_of_motion_noiseless}
    \left[ Z i\partial_t - r + i\kappa \right] b_c(t) - \lambda |b_c(t)|^2 b_c(t) = 0\,.
\end{align}
The term $iD|b_q|^2$ corresponds to averaging over a complex Gaussian noise $\xi(t)$ with correlator
\begin{align}
    \langle \xi^*(t) \xi(t') \rangle = 2D \, \delta(t - t').
\label{eq:noise_correlator}
\end{align}
Including noise, the effective Langevin equation reads
\begin{align}
    Z \,i \dot{b}_c(t) = \left[ r - i\kappa \right] b_c(t) + \lambda |b_c(t)|^2 b_c(t) + \xi(t),
\label{eq:langevin_equation}
\end{align}
with $\xi$ as above. This is the canonical single-mode laser Langevin equation in Keldysh language:
\begin{itemize}
\item $r$ is the distance from threshold,
\item $\kappa$ is the linear gain/damping,
\item $\lambda$ saturates the growth,
\item $D$ sets the strength of noise injected by the beam.
\end{itemize}
The global phase symmetry $b_c \to b_c e^{i\theta}$ is inherited from the $U(1)$ invariance of Preparata's Lagrangian and is spontaneously broken in the lasing phase, where $\langle b_c\rangle \neq 0$ selects one phase out of a continuum. This pattern closely parallels the scenario discussed in Ref.~\cite{DelGiudiceVitiello2006}, where the electromagnetic field in a many-body system develops a nonvanishing order parameter and a massless phase mode, leading to extended coherent domains. In our case, the FEL order parameter is the coherent electromagnetic amplitude, and the lasing transition realizes the same $U(1)$ symmetry breaking in a driven, non-equilibrium setting.

\subsubsection{Stationary solutions and the order parameter}

Stationary solutions of Eq.~\eqref{eq:langevin_equation} (ignoring noise) satisfy
\begin{align}
0 = (r - i\kappa) b_c + \lambda |b_c|^2 b_c.
\label{eq:stationary_condition}
\end{align}
Besides the trivial solution $b_c=0$, non-trivial solutions obey
\begin{align}
r - i\kappa + \lambda |b_c|^2 = 0.
\label{eq:nontrivial_condition}
\end{align}
Writing $\lambda = \lambda_R + i\lambda_I$ and separating real and imaginary parts yields
\begin{align}
r + \lambda_R |b_c|^2 &= 0, \label{eq:real_part} \\
-\kappa + \lambda_I |b_c|^2 &= 0. \label{eq:imag_part}
\end{align}
For a choice of phase where $\lambda$ is predominantly real, the amplitude scales as
\begin{align}
|b_c|^2 \approx -\frac{r}{\text{Re}\, \lambda},
\label{eq:order_parameter_amplitude}
\end{align}
so that $|b_c|\propto r^{1/2}$ for $r>0$. The threshold $r=0$ thus marks a continuous transition from a normal phase ($b_c=0$) to a lasing phase with finite coherent field.

\subsection{Universality class and critical fluctuations}

After rescaling to a frame where the purely imaginary part of the linear coefficient is absorbed, Eq.~\eqref{eq:langevin_equation} takes the standard laser form
\begin{align}
    \dot{a}(t) = \left[ \alpha - \beta |a(t)|^2 \right] a(t) + \zeta(t),\label{eq:canonical_laser} \\
    \langle \zeta^*(t) \zeta(t') \rangle = 2D_{\text{las}} \, \delta(t - t'),\nonumber
\end{align}
with $a(t)$ a rescaled version of $b_c(t)$ and $(\alpha,\beta,D_{\text{las}})$ linear combinations of $(r,\kappa,\lambda,D,Z)$. The FEL threshold in the Preparata model thus belongs to the laser universality class: a zero-dimensional complex order parameter with $U(1)$ symmetry, driven-dissipative dynamics, and Markovian noise. Critical exponents are mean-field: $|b_c|\sim r^{1/2}$, correlation time $\tau\sim |r|^{-1}$, susceptibility $\chi\sim |r|^{-1}$.

Restoring the noise and linearizing Eq.~\eqref{eq:langevin_equation} near $b_c=0$,
\begin{align}
Z i\partial_t b_c(t) = r \, b_c(t) + \xi(t),
\label{eq:linear_langevin}
\end{align}
one finds an Ornstein–Uhlenbeck process with stationary variance
\begin{align}
\langle |b_c|^2 \rangle \sim \frac{D}{r},
\label{eq:critical_fluctuations}
\end{align}
and relaxation time
\begin{align}
\tau \sim \frac{1}{|r|}.
\label{eq:correlation_time}
\end{align}
As $r\to 0$, fluctuations and correlation times diverge, and their amplitudes are fixed by the microscopic noise kernel $\Sigma^K(\omega)$, i.e. by beam current correlators.

\begin{figure}[h]
\centering
\begin{tikzpicture}[scale=1.5]

\fill[blue!12] (-2.3,-0.3) -- (-2.3,3) -- (0,3) -- (0,-0.3) -- cycle;
\fill[red!12] (0,-0.3) -- (0,3) -- (2.8,3) -- (2.8,-0.3) -- cycle;

\draw[thick, ->] (-2.3,0) -- (2.9,0) node[right, font=\large] {$r$};
\draw[thick, ->] (0,-0.3) -- (0,3.1) node[above, font=\large] {$|b_c|$};

\draw[blue, very thick] (-2.3,0) -- (-0.02,0);
\draw[red, very thick, domain=0.02:2.8, samples=120] plot (\x, {sqrt(\x)});

\draw[gray, dashed, very thick] (0,-0.3) -- (0,3.1);
\filldraw[black] (0,0) circle (0.12);

\node[font=\Large\bfseries\color{blue}] at (-1.1, -0.8) {Normal phase};
\node[font=\Large\bfseries\color{red}] at (1.4, -0.8) {Lasing phase};

\node[font=\large\color{blue}] at (-1.2, 0.25) {$|b_c|=0$};
\node[font=\large\color{red}] at (1.6, 1.7) {$|b_c| \propto \sqrt{r}$};

\node[font=\normalsize, black] at (0.55, 2.95) {$r=0$};

\end{tikzpicture}
\caption{Order parameter $|b_c|$ as a function of distance from threshold $r = Z\Delta\omega$. 
For $r < 0$ the trivial state $b_c = 0$ is stable (normal phase). At $r = 0$ the system 
undergoes a continuous phase transition. For $r > 0$ a finite-amplitude coherent state 
emerges with $|b_c| \propto \sqrt{r}$ (lasing phase).}
\label{fig:phase_transition}
\end{figure}

\section{Comparison with standard FEL theory}

We now relate the Keldysh formulation to standard FEL theory, in particular to the classical dispersion relation, the Pierce parameter, and Madey's gain formula.

\subsection{Classical FEL dispersion relation from the self-energy}

In classical one-dimensional FEL theory, the dispersion relation typically reads
\begin{align}
\left( \omega - \omega_{\text{res}} \right) = \text{const.} \times \int dp \, \frac{\partial f/\partial p}{p - p_{\text{res}}},
\label{eq:classical_dispersion}
\end{align}
where $f(p)$ is the beam distribution, and the integral is understood as a principal value.

In our formalism, the analogous relation is $\Gamma^R(\omega)=0$, with
\begin{align}
\Gamma^R(\omega) = \frac{N}{2\pi}(\omega - \omega_\eta) - \Sigma^R(\omega),
\label{eq:dressed_propagator_recall}
\end{align}
and
\begin{align}
\Sigma^R(\omega) = N\eta \sum_m (n_m - n_{m+1}) \left[ \mathcal{P}\frac{1}{\omega - \Omega_m} - i\pi\delta(\omega - \Omega_m) \right].
\label{eq:sigma_R_recall}
\end{align}
Taking the continuum limit $\sum_m\to\int dm$, introducing a momentum-like variable $p=m/\sqrt{\eta}$ and a distribution $f(p)$ such that $n(m)=\sqrt{\eta}f(p)$, one finds
\begin{align}
n_m - n_{m+1} \approx -\sqrt{\eta}\,\frac{\partial f}{\partial p}, \qquad \Omega_m \approx \frac{p}{\sqrt{\eta}}.
\label{eq:population_difference_continuum}
\end{align}
Substituting into Eq.~\eqref{eq:sigma_R_recall} leads to
\begin{align}
    \Sigma^R(\omega) \approx -N\eta^{3/2} \int dp \, \frac{\partial f/\partial p}{\omega - p/\sqrt{\eta}},
\label{eq:sigma_R_continuum}
\end{align}
and the threshold condition $\Gamma^R(\omega)=0$ becomes
\begin{align}
\frac{\omega - \omega_\eta}{2\pi} = -\eta^{3/2} \int dp \, \frac{\partial f/\partial p}{\omega - p/\sqrt{\eta}},
\label{eq:classical_dispersion_recovered}
\end{align}
which has the same structure as Eq.~\eqref{eq:classical_dispersion}. The appearance of $\partial f/\partial p$ originates directly from the population difference $(n_m-n_{m+1})$ and provides a microscopic derivation of the central ingredient entering classical FEL gain formulas.

\subsection{Connection to the Pierce parameter}

For a cold beam sharply peaked at $m_0$, taking $n_m\simeq\delta_{m,m_0}$, the self-energy reduces to
\begin{align}
    \Sigma^R(\omega)& \approx N\eta \left[ \frac{1}{\omega - \Omega_{m_0}} - \frac{1}{\omega - \Omega_{m_0+1}} \right] 
    \nonumber\\
    &= N\eta \, \frac{\Omega_{m_0+1} - \Omega_{m_0}}{(\omega - \Omega_{m_0})(\omega - \Omega_{m_0+1})}.
\label{eq:cold_beam_self_energy}
\end{align}
Close to $\Omega_0:=\Omega_{m_0}$ this yields $\Sigma^R(\omega)\propto (\omega-\Omega_0)^{-2}$, and the condition $\Gamma^R(\omega)=0$ becomes a cubic equation in $\delta\omega=\omega-\Omega_0$, analogous to the Pierce cubic dispersion relation. The effective Pierce parameter $\rho_{\text{eff}}$ can be identified from the coefficient of this cubic and scales as
\begin{align}
\rho_{\text{eff}}^3 \sim \frac{N\eta}{m_0},
\label{eq:pierce_parameter_identification}
\end{align}
recovering the expected dependence on coupling strength and beam energy. The growth rate then scales as $\sim \rho_{\text{eff}}\omega_0$, as in the classical Pierce model.

\subsection{Madey gain formula and energy spread}

Madey's theorem relates the gain to the derivative $\partial f/\partial \varepsilon$ at resonance, and predicts suppression of gain as energy spread increases. In our formulation, gain is governed by $\text{Im}\,\Sigma^R(\omega)$, which for the Gaussian beam reads
\begin{align}
\text{Im}\, \Sigma^R(\omega) \approx -N\sqrt{\pi} \, \frac{y}{\sigma_\Omega^2} e^{-y^2}.
\label{eq:gain_formula_recall}
\end{align}
Relating $\sigma_\Omega$ to the energy spread $\sigma_\varepsilon$ via the beam dispersion shows that gain is suppressed as $1/\sigma_\varepsilon^2$, in agreement with classical FEL scaling laws. Maximum gain occurs at $y\sim O(1)$, i.e. slightly away from the distribution center, again matching standard FEL results for beams with finite spread.

In conclusion, the Keldysh self-energy formulation reproduces the classical FEL dispersion relation, Pierce scaling, and Madey-type gain suppression when appropriate limits are taken, while at the same time providing a microscopic and unified description of dispersion, gain, and noise.

\section{Conclusions}

We have developed a non-equilibrium quantum field theory of the free-electron laser based on the Schwinger-Keldysh formalism and the Preparata model. Starting from the microscopic Lagrangian for a relativistic electron beam coupled to a single radiation mode, we constructed a real-time functional integral, integrated out the electrons, and obtained an effective action for the FEL mode where dispersion, gain, and noise are all encoded in a single object: the electronic self-energy, expressed in terms of current correlation functions of the beam.

For a stationary Gaussian beam we derived closed analytic expressions for all components of the self-energy in terms of beam parameters (current, energy spread, detuning), involving the Dawson integral. These results provide a compact and transparent description of frequency pulling, gain and damping, and noise spectra.

At low frequencies, the theory reduces to a Landau-Ginzburg-Keldysh description with effective mass $r$, growth rate $\kappa$, nonlinearity $\lambda$, and noise strength $D$ fully determined by the microscopic self-energies. The FEL threshold appears as a continuous non-equilibrium phase transition in the laser universality class, with the coherent radiation amplitude as order parameter and mean-field critical behavior. Critical fluctuations near threshold are controlled by the microscopic noise kernel $\Sigma^K(\omega)$ rather than by phenomenological stochastic terms.

The Keldysh formulation recovers standard FEL results—the classical dispersion relation, Pierce cubic scaling, and Madey gain suppression—while providing a unified and systematic framework for including fluctuations and response on equal footing. This perspective naturally connects FEL physics to broader developments in non-equilibrium quantum field theory and driven–dissipative systems, and offers a route to extend FEL theory to regimes where quantum noise, non-Gaussian fluctuations, or multimode and spatial effects become important.

\appendix
\section{From contour fields to the Keldysh action in the $(c,q)$ basis}
\label{app:keldysh-rotation}

In this Appendix, we show explicitly how the Keldysh action in the $(c,q)$ basis is obtained from the contour
representation in Eq.~\eqref{def:keldish-action} by performing the rotation \eqref{def:rotation-xi}–\eqref{def:rotation-b}
and expanding in powers of the quantum fields. This makes transparent the origin of the three structural pieces of
the effective action: free matter, free radiation, and their interaction.

\subsection{General structure on the Keldysh contour}

Let us split the microscopic action $S[\xi,b]$ in Eq.~\eqref{eq:FEL-L-rescaled} into matter, radiation and
interaction contributions,
\begin{equation}
  S[\xi,b] = S_M[\xi] + S_B[b] + S_{\text{int}}[\xi,b].
\end{equation}
On the Keldysh contour, the action difference entering \eqref{def:keldish-action} is
\begin{equation}
\begin{split}
  S_K[\xi_+,\xi_-;b_+,b_-]
  &= S[\xi_+,b_+] - S[\xi_-,b_-]
  \\
  &= S_{M,K} + S_{B,K} + S_{\text{int},K},
\end{split}
\end{equation}
with
\begin{align}
  S_{M,K} &= S_M[\xi_+] - S_M[\xi_-],\\
  S_{B,K} &= S_B[b_+] - S_B[b_-],\\
  S_{\text{int},K} &= S_{\text{int}}[\xi_+,b_+] - S_{\text{int}}[\xi_-,b_-].
\end{align}
We now treat each term in turn, using the classical/quantum fields defined in
Eqs.~\eqref{def:rotation-xi}–\eqref{def:rotation-b},
\begin{equation}
  \xi_\pm = \xi_c \pm \frac{1}{2}\xi_q,
  \qquad
  b_\pm   = b_c \pm \frac{1}{2}b_q,
\end{equation}
and the analogous relations for the adjoint fields.

Throughout we work in the regime of small quantum fields, retaining terms at most linear in $\xi_q$ and $b_q$ in the
action. This is sufficient for Gaussian fluctuations and linear response and leads directly to the Landau–Ginzburg–
Keldysh structure used in the main text.

\subsection{Free matter contribution}

The free matter part of the microscopic action has the generic quadratic form
\begin{equation}
  S_M[\xi]
  = N\int_{t_0}^{t_f} dt \int_0^{2\pi} d\phi\,
    \xi^\dagger(\phi,t)\,\mathcal{L}\,\xi(\phi,t),
\end{equation}
where $\mathcal{L}$ is the single-particle operator appearing in Eq.~\eqref{eq:FEL-L-rescaled} (in our case
$\mathcal{L} = i\partial_t + i\dot{\phi}\partial_\phi - \varepsilon(-i\partial_\phi)$ up to a surface term). On the
contour, the Keldysh combination is
\begin{equation}
  S_{M,K}
  = N\int_{t_0}^{t_f} dt \int_0^{2\pi} d\phi\,
    \big[\xi_+^\dagger \mathcal{L}\,\xi_+ - \xi_-^\dagger \mathcal{L}\,\xi_-\big].
  \label{app:SMK-start}
\end{equation}
We now insert $\xi_\pm = \xi_c \pm \xi_q/2$. For the $+$ branch,
\begin{align}
  \xi_+^\dagger \mathcal{L}\,\xi_+
  &= \big(\xi_c^\dagger + \tfrac{1}{2}\xi_q^\dagger\big)\,
     \mathcal{L}\,
     \big(\xi_c + \tfrac{1}{2}\xi_q\big)\nonumber\\
  &= \xi_c^\dagger \mathcal{L}\,\xi_c
   + \frac{1}{2}\xi_q^\dagger \mathcal{L}\,\xi_c
   + \frac{1}{2}\xi_c^\dagger \mathcal{L}\,\xi_q
   + \frac{1}{4}\xi_q^\dagger \mathcal{L}\,\xi_q.
\end{align}
For the $-$ branch,
\begin{align}
  \xi_-^\dagger \mathcal{L}\,\xi_-
  &= \big(\xi_c^\dagger - \tfrac{1}{2}\xi_q^\dagger\big)\,
     \mathcal{L}\,
     \big(\xi_c - \tfrac{1}{2}\xi_q\big)\nonumber\\
  &= \xi_c^\dagger \mathcal{L}\,\xi_c
   - \frac{1}{2}\xi_q^\dagger \mathcal{L}\,\xi_c
   - \frac{1}{2}\xi_c^\dagger \mathcal{L}\,\xi_q
   + \frac{1}{4}\xi_q^\dagger \mathcal{L}\,\xi_q.
\end{align}
Taking the difference, the purely classical terms and the quadratic quantum terms cancel exactly:
\begin{align}
  \xi_+^\dagger \mathcal{L}\,\xi_+ - \xi_-^\dagger \mathcal{L}\,\xi_-
  &= \big(\tfrac{1}{2}+\tfrac{1}{2}\big)\xi_q^\dagger \mathcal{L}\,\xi_c
   + \big(\tfrac{1}{2}+\tfrac{1}{2}\big)\xi_c^\dagger \mathcal{L}\,\xi_q\nonumber\\
  &= \xi_q^\dagger \mathcal{L}\,\xi_c + \xi_c^\dagger \mathcal{L}\,\xi_q.
\end{align}
Substituting back into \eqref{app:SMK-start}, we obtain
\begin{align}
\begin{split}
  S_{M,K}[\xi_c,\xi_q]
  = N\int_{t_0}^{t_f} dt \int_0^{2\pi} d\phi\,\big[\xi_q^\dagger&(\phi,t)\,\mathcal{L}\,\xi_c(\phi,t)\\
  &+\xi_c^\dagger(\phi,t)\,\mathcal{L}\,\xi_q(\phi,t)
    \big],\nonumber
\end{split}
\end{align}
which is the expression used in the main text. The $(c,q)$ rotation has thus converted the difference of two quadratic
contour actions into a bilinear form in which the quantum field $\xi_q$ plays the role of a Lagrange multiplier
enforcing the classical equations of motion for $\xi_c$.

\subsection{Free radiation contribution}

The free part of the radiation action is also quadratic and can be written as
\begin{equation}
  S_B[b]
  = \int_{t_0}^{t_f} dt\, b^\dagger(t)\,\mathcal{L}_B\,b(t),
\end{equation}
where $\mathcal{L}_B$ is the single-mode operator (in our case, $\mathcal{L}_B = N( i\partial_t - \omega_\eta)$ after
rescaling). The corresponding Keldysh combination is
\begin{equation}
  S_{B,K}
  = \int_{t_0}^{t_f} dt\,
    \big[b_+^\dagger \mathcal{L}_B b_+ - b_-^\dagger \mathcal{L}_B b_-\big].
\end{equation}
Repeating the same algebra as for the matter sector, with
$b_\pm = b_c \pm b_q/2$, we obtain
\begin{equation}
  S_{B,K}[b_c,b_q]
  = \int_{t_0}^{t_f} dt\,
    \big[
      b_q^\dagger(t)\,\mathcal{L}_B\,b_c(t)
      + b_c^\dagger(t)\,\mathcal{L}_B\,b_q(t)
    \big].
\end{equation}
This is the quadratic part of the Keldysh action for the FEL mode that leads, after integrating out the matter fields,
to the dressed propagator $\Gamma^{R,A}$ used in the main text.

\subsection{Interaction contribution and expansion in quantum fields}

The interaction term couples the FEL mode to the electronic current at the resonant wavevector. For definiteness we
take the interaction in the form
\begin{equation}
\begin{split}
  S_{\text{int}}[\xi,b]
  = \eta\int_{t_0}^{t_f}\!\!\! dt \int_0^{2\pi}\!\!\! d\phi\,
    \big[
     & b(t)\,\xi^\dagger(\phi,t)\,e^{-i\phi}\,\xi(\phi,t)
      \\
      &+ b^\dagger(t)\,\xi^\dagger(\phi,t)\,e^{i\phi}\,\xi(\phi,t)
    \big],\nonumber
\end{split}
\end{equation}
where the combination $\xi^\dagger e^{-i\phi}\xi$ is the mode of the electronic current that couples to the FEL field. On the contour, the Keldysh version is
\begin{align}
  S_{\text{int},K}
  &= S_{\text{int}}[\xi_+,b_+] - S_{\text{int}}[\xi_-,b_-]\nonumber\\
  &= \eta\int_{t_0}^{t_f} dt \int_0^{2\pi} d\phi\,
     \Big\{
       b_+\,J_+^\dagger + b_+^\dagger J_+
         \label{app:SintK-start}\\
       &\qquad\qquad\qquad\qquad\qquad\qquad- b_-\,J_-^\dagger - b_-^\dagger J_-
     \Big\},\nonumber
\end{align}
where we have defined the current mode
\begin{equation}
  J_\pm(\phi,t) := \xi_\pm^\dagger(\phi,t)\,e^{-i\phi}\,\xi_\pm(\phi,t).
\end{equation}
We now express $b_\pm$ and $J_\pm$ in terms of classical and quantum fields. For the field,
\begin{equation}
  b_\pm = b_c \pm \frac{1}{2}b_q.
\end{equation}
For the current, we expand $J_\pm$ to leading order in $\xi_q$,
\begin{align}
  J_+ &= (\xi_c^\dagger + \tfrac{1}{2}\xi_q^\dagger)\,e^{-i\phi}\,(\xi_c + \tfrac{1}{2}\xi_q)\nonumber\\
      &\qquad\qquad= J_c + \frac{1}{2}J_q + \mathcal{O}(\xi_q^2),\\[4pt]
  J_- &= (\xi_c^\dagger - \tfrac{1}{2}\xi_q^\dagger)\,e^{-i\phi}\,(\xi_c - \tfrac{1}{2}\xi_q)\nonumber\\
      &\qquad\qquad= J_c - \frac{1}{2}J_q + \mathcal{O}(\xi_q^2),
\end{align}
where we introduced
\begin{align*}
  J_c(\phi,t) &:= \xi_c^\dagger(\phi,t)\,e^{-i\phi}\,\xi_c(\phi,t),\\
  J_q(\phi,t) &:= \xi_q^\dagger(\phi,t)\,e^{-i\phi}\,\xi_c(\phi,t)
                + \xi_c^\dagger(\phi,t)\,e^{-i\phi}\,\xi_q(\phi,t).
\end{align*}
Substituting these expressions into \eqref{app:SintK-start} yields, up to quadratic order in the quantum fields,
\begin{equation}
\begin{split}
  S_{\text{int},K}
  \simeq \eta\int_{t_0}^{t_f} dt \int_0^{2\pi} &d\phi\,
     \Big[
       (b_c+\tfrac{1}{2}b_q)(J_c^\dagger+\tfrac{1}{2}J_q^\dagger)\\
       &+(b_c^\dagger+\tfrac{1}{2}b_q^\dagger)(J_c+\tfrac{1}{2}J_q)\\
     &- (b_c-\tfrac{1}{2}b_q)(J_c^\dagger-\tfrac{1}{2}J_q^\dagger)\\
     &- (b_c^\dagger-\tfrac{1}{2}b_q^\dagger)(J_c-\tfrac{1}{2}J_q)
     \Big].
\end{split}
\end{equation}
Expanding and collecting terms, the contributions purely quadratic in classical fields cancel, and to leading order in
the quantum fields we obtain
\begin{align}
  S_{\text{int},K}
  \simeq \eta\int_{t_0}^{t_f}\!\!\! dt \int_0^{2\pi} \!\!\!\!d\phi\,
     \Big[
       b_q\,J_c^\dagger + b_c\,J_q^\dagger
       &+ b_q^\dagger J_c + b_c^\dagger J_q
     \Big]
     \nonumber\\
     &+ \mathcal{O}(b_q J_q,\,\xi_q^2).
\end{align}
The terms of order $b_q J_q$ and quadratic in $\xi_q$ are higher order in the quantum fields and are neglected in the
Gaussian approximation used in the main text. In this approximation, the interaction contribution reduces to
\begin{equation*}
  S_{\text{int},K}
  \simeq \eta\int_{t_0}^{t_f} dt \int_0^{2\pi} d\phi\,
     \Big[
       b_q\,J_c^\dagger + b_c\,J_q^\dagger
       + b_q^\dagger J_c + b_c^\dagger J_q
     \Big],
\end{equation*}
which is the starting point for integrating out the matter fields and deriving the self-energy structure of the FEL
mode.

\section{Derivation of the electronic self-energy via Wick's theorem}
\label{app:wick}

Here, we provide a detailed derivation of the quadratic influence functional and of the electronic self-energies $\Sigma^{R,A,K}$ using Wick's theorem. Although standard in Keldysh formalism, we present it here pedagogically for the benefit of readers unfamiliar with Keldysh techniques.

\subsection{Cumulant expansion}

The radiation field amplitude $b$ is small near the FEL threshold; thus, then we can expand the influence functional in powers of $b$. Taking the logarithm of both sides of Eq.~\eqref{def:action-IF}, we have
\begin{equation}
    S_{\text{IF}}[b_c, b_q] = -i \ln \left\langle e^{i S_{\text{int},K}} \right\rangle_0,
    \label{eq:log-expand}
\end{equation}
where $\langle \cdots \rangle_0$ denotes the path integral average with respect to the free action $S_{M,K}$ (no coupling to $b$). Expanding the exponential and the logarithm in powers of $S_{\text{int},K}$, the cumulant expansion gives
\begin{align}
    S_{\text{IF}} &= -i \ln \left[ 1 + i \langle S_{\text{int},K} \rangle_0 + \frac{i^2}{2!} \langle S_{\text{int},K}^2 \rangle_0 + \cdots \right] \nonumber\\
    &= \langle S_{\text{int},K} \rangle_0 - \frac{1}{2} \langle S_{\text{int},K}^2 \rangle_{0,c} + \mathcal{O}(b^3),
    \label{eq:cumulant-expansion}
\end{align}
where $\langle \cdots \rangle_{0,c}$ denotes the \emph{connected} correlator:
\begin{equation}
    \langle S_{\text{int},K}^2 \rangle_{0,c} = \langle S_{\text{int},K}^2 \rangle_0 - \langle S_{\text{int},K} \rangle_0^2.
\end{equation}

\paragraph{First-order term.}
The first-order contribution is
\begin{align}
    \langle S_{\text{int},K} \rangle_0 &= -iN\eta^{1/2} \int dt\; \Big[ b_c^\dagger(t) \langle J_q(t) \rangle_0 - b_c(t) \langle J_q^\dagger(t) \rangle_0 \nonumber\\
    &\qquad\qquad\qquad+ b_q^\dagger(t) \langle J_c(t) \rangle_0 - b_q(t) \langle J_c^\dagger(t) \rangle_0 \Big].
\end{align}
For a beam state with no initial macroscopic density modulation at the FEL wavevector, we have $\langle J_c(t) \rangle_0 = 0$ and $\langle J_q(t) \rangle_0 = 0$. Therefore, the first-order term vanishes:
\begin{equation}
    \langle S_{\text{int},K} \rangle_0 = 0.
    \label{eq:first-order-zero}
\end{equation}

\paragraph{Second-order term.}
The second-order contribution is
\begin{align}
    S_{\text{IF}}^{(2)} &= -\frac{1}{2} \langle S_{\text{int},K}^2 \rangle_{0,c} \nonumber\\
    &= -\frac{1}{2} \langle S_{\text{int},K}^2 \rangle_0,
    \label{eq:second-order-start}
\end{align}
where we used $\langle S_{\text{int},K} \rangle_0 = 0$ in the last step. Squaring $S_{\text{int},K}$, we obtain
\begin{widetext}
\begin{align}
    S_{\text{int},K}^2 = (N\eta) \int dt \int dt' \, \Big[ b_c^\dagger(t) J_q(t) &- b_c(t) J_q^\dagger(t) + b_q^\dagger(t) J_c(t) - b_q(t) J_c^\dagger(t) \Big] \nonumber\\
    &\qquad\qquad\qquad\times \Big[ b_c^\dagger(t') J_q(t') - b_c(t') J_q^\dagger(t') + b_q^\dagger(t') J_c(t') - b_q(t') J_c^\dagger(t') \Big].
    \label{eq:Sint-squared}
\end{align}
\end{widetext}
Expanding this product gives 16 terms. We organize them according to which components of $b$ appear.

\subsection{Application of Wick's theorem}

We now use Wick's theorem to evaluate the expectation values. For Gaussian states (which include the free electron action $S_{M,K}$), Wick's theorem states that the expectation value of a product of operators can be expressed as a sum over all possible pairwise contractions~\cite{kamenev}. For our purposes, the key contractions are:
\begin{align}
    \langle J_c(t) J_c^\dagger(t') \rangle_0 &\neq 0, \\
    \langle J_q(t) J_q^\dagger(t') \rangle_0 &\neq 0, \\
    \langle J_c(t) J_q^\dagger(t') \rangle_0 &= 0, \\
    \langle J_q(t) J_c^\dagger(t') \rangle_0 &= 0.
\end{align}
The last two relations follow from the fact that $J_c$ contains only products $\xi_c^\dagger \xi_c$, while $J_q$ contains $\xi_c^\dagger \xi_q + \xi_q^\dagger \xi_c$; there are no contractions that connect $c$ and $q$ fields in a free Gaussian theory.

\paragraph{Terms quadratic in $b_c$.}
Consider the terms in Eq.~\eqref{eq:Sint-squared} that are proportional to $b_c^\dagger(t) b_c(t')$:
\begin{equation}
    (N\eta) \int dt \int dt' \, b_c^\dagger(t) b_c(t') \times \Big[ J_q(t) \otimes (-J_q^\dagger(t')) \Big].
\end{equation}
Taking the expectation value:
\begin{equation}
    -(N\eta) \int dt \int dt' \, b_c^\dagger(t) b_c(t') \langle J_q(t) J_q^\dagger(t') \rangle_0.
    \label{eq:bc-bc-term}
\end{equation}
Similarly, the $b_c(t) b_c^\dagger(t')$ terms give:
\begin{equation}
    -(N\eta) \int dt \int dt' \, b_c(t) b_c^\dagger(t') \langle J_q^\dagger(t) J_q(t') \rangle_0.
    \label{eq:bc-bc-herm}
\end{equation}

\paragraph{Terms quadratic in $b_q$.}
Similarly, the terms proportional to $b_q^\dagger(t) b_q(t')$ give:
\begin{equation}
    -(N\eta) \int dt \int dt' \, b_q^\dagger(t) b_q(t') \langle J_c(t) J_c^\dagger(t') \rangle_0,
    \label{eq:bq-bq-term}
\end{equation}
and the $b_q(t) b_q^\dagger(t')$ terms give:
\begin{equation}
    -(N\eta) \int dt \int dt' \, b_q(t) b_q^\dagger(t') \langle J_c^\dagger(t) J_c(t') \rangle_0.
    \label{eq:bq-bq-herm}
\end{equation}

\paragraph{Mixed terms $b_c b_q$.}
The cross terms proportional to $b_c^\dagger(t) b_q(t')$ give:
\begin{equation}
    -(N\eta) \int dt \int dt' \, b_c^\dagger(t) b_q(t') \langle J_q(t) J_c^\dagger(t') \rangle_0 = 0,
\end{equation}
which vanishes because $J_q$ and $J_c$ do not contract in a free theory. Similarly, all other $b_c \leftrightarrow b_q$ cross terms vanish.

\subsection{Reorganizing into Keldysh matrix form}

We now reorganize the non-vanishing terms into the standard Keldysh $2\times2$ matrix structure. The second-order influence functional is
\begin{align}
    S_{\text{IF}}^{(2)} &= -\frac{1}{2}(N\eta) \int dt \int dt' \, \Big\{ b_c^\dagger(t) b_c(t') \langle J_q(t) J_q^\dagger(t') \rangle_0 \nonumber\\
    &\qquad+ b_c(t) b_c^\dagger(t') \langle J_q^\dagger(t) J_q(t') \rangle_0 \nonumber\\
    &\qquad+ b_q^\dagger(t) b_q(t') \langle J_c(t) J_c^\dagger(t') \rangle_0 \nonumber\\
    &\qquad+ b_q(t) b_q^\dagger(t') \langle J_c^\dagger(t) J_c(t') \rangle_0 \Big\}.
    \label{eq:SIF2-collected}
\end{align}

\paragraph{Introducing retarded, advanced, and Keldysh correlators.}
We now define the standard decomposition of the two-time correlators into retarded, advanced, and Keldysh components. For the classical current $J_c$:
\begin{align}
    C^R_{cc}(t,t') &= -i\theta(t-t') \langle [J_c(t), J_c^\dagger(t')] \rangle_0, \\
    C^A_{cc}(t,t') &= +i\theta(t'-t) \langle [J_c(t), J_c^\dagger(t')] \rangle_0, \\
    C^K_{cc}(t,t') &= -i \langle \{J_c(t), J_c^\dagger(t')\} \rangle_0,
\end{align}
where $[\cdot,\cdot]$ is the commutator and $\{\cdot,\cdot\}$ is the anticommutator. These satisfy the identity
\begin{equation}
    \langle J_c(t) J_c^\dagger(t') \rangle_0 = C^R_{cc}(t,t') + \frac{i}{2} C^K_{cc}(t,t').
    \label{eq:decomp-Jc}
\end{equation}
Similarly for the quantum current $J_q$:
\begin{align}
    C^R_{qq}(t,t') &= -i\theta(t-t') \langle [J_q(t), J_q^\dagger(t')] \rangle_0, \\
    C^A_{qq}(t,t') &= +i\theta(t'-t) \langle [J_q(t), J_q^\dagger(t')] \rangle_0, \\
    C^K_{qq}(t,t') &= -i \langle \{J_q(t), J_q^\dagger(t')\} \rangle_0.
\end{align}

\paragraph{Key observation: Keldysh rotation.}
A crucial property of the Keldysh formalism is that the current operators in the $(c,q)$ basis are related to the $(+,-)$ basis by~\cite{kamenev}
\begin{equation}
    J = \frac{J_+ + J_-}{2}, \qquad J_q = J_+ - J_-.
\end{equation}
Using the fact that $J_\pm$ commute with themselves (they act on different branches of the Schwinger-Keldysh contour), one can show that
\begin{align}
    C^R_{qq}(t,t') &= C^R_{cc}(t,t'), \\
    C^K_{qq}(t,t') &= -C^K_{cc}(t,t').
\end{align}
This is a general property of the Keldysh rotation.

\subsection{Final result: the self-energy matrix}

Using the decomposition in Eq.~\eqref{eq:decomp-Jc} and the Keldysh rotation properties, we can rewrite Eq.~\eqref{eq:SIF2-collected} as
\begin{align}
    S_{\text{IF}}^{(2)} &= -\int dt \int dt' \, \Big\{ b_c^\dagger(t) \Sigma^R(t,t') b_c(t') \nonumber\\
    &\qquad+ b_q^\dagger(t) \Sigma^A(t,t') b_c(t') + b_c^\dagger(t) \Sigma^A(t,t') b_q(t') \nonumber\\
    &\qquad+ \frac{1}{2} \left[ b_q^\dagger(t) \Sigma^K(t,t') b_q(t') + b_q(t) \Sigma^K(t,t') b_q^\dagger(t') \right] \Big\},
\end{align}
where we have defined the electronic self-energies:
\begin{align}
    \Sigma^R(t,t') &= -iN\eta \, \theta(t-t') \langle [J(t), J^\dagger(t')] \rangle_0, \\
    \Sigma^A(t,t') &= +iN\eta \, \theta(t'-t) \langle [J(t), J^\dagger(t')] \rangle_0, \\
    \Sigma^K(t,t') &= -iN\eta \langle \{J(t), J^\dagger(t')\} \rangle_0.
\end{align}
Here we have used $J = J_c$ in the expectation values (the quantum component $J_q$ does not contribute to expectation values in the free theory).

This can be written compactly in the standard Keldysh matrix form:
\begin{widetext}
\begin{equation}
    S_{\text{IF}}^{(2)}[b_c, b_q] = -\int dt \, dt' \begin{pmatrix} b_q^\dagger(t) & b_c^\dagger(t) \end{pmatrix} \begin{pmatrix} 0 & \Sigma^A(t,t') \\ \Sigma^R(t,t') & \Sigma^K(t,t') \end{pmatrix} \begin{pmatrix} b_c(t') \\ b_q(t') \end{pmatrix}.
    \label{eq:influence_functional_final}
\end{equation}
\end{widetext}

This completes the derivation of the electronic self-energy from Wick's theorem.

\section{Low-frequency expansion and effective time-domain action}
\label{app:low_frequency}

We now derive the effective time-domain action~\eqref{eq:effective_action_time} from the frequency-space form~\eqref{eq:effective_action_frequency} by performing a systematic low-frequency expansion around the FEL threshold. This procedure is standard in the derivation of effective theories for critical phenomena~\cite{cross_hohenberg,tauber2014critical}.

We begin by defining the full retarded inverse propagator (also called the dressed propagator)
\begin{equation}
    \Gamma^R(\omega) = [G_0^R(\omega)]^{-1} - \Sigma^R(\omega) = \frac{N}{2\pi}(\omega - \omega_\eta) - \Sigma^R(\omega).
    \label{eq:GammaR_def_app}
\end{equation}
Similarly, the advanced propagator is
\begin{equation}
    \Gamma^A(\omega) = [G_0^A(\omega)]^{-1} - \Sigma^A(\omega) = [\Gamma^R(\omega)]^*.
    \label{eq:GammaA_def_app}
\end{equation}
The effective action becomes
\begin{align}
    S_{\text{eff}}^{(2)} &= \int \frac{d\omega}{2\pi} \Big\{ b_q^\dagger(\omega) \Gamma^R(\omega) b_c(\omega) \nonumber\\
    &\quad+ b_c^\dagger(\omega) \Gamma^A(\omega) b_q(\omega) - b_c^\dagger(\omega) \Sigma^K(\omega) b_c(\omega) \Big\}.
    \label{eq:Seff_Gamma}
\end{align}

Near the lasing threshold, the relevant dynamics occurs at frequencies $\omega$ close to the beam resonance. We work in a rotating frame where $\omega = 0$ corresponds to the (shifted) resonance frequency. We then expand $\Gamma^R(\omega)$ around $\omega = 0$:
\begin{equation}
    \Gamma^R(\omega) = \Gamma^R(0) + \omega \frac{\partial \Gamma^R}{\partial \omega}\bigg|_{\omega=0} + \mathcal{O}(\omega^2).
    \label{eq:GammaR_expansion}
\end{equation}
We define the field renormalization factor
\begin{equation}
    Z^{-1} = \frac{\partial \Gamma^R(\omega)}{\partial \omega}\bigg|_{\omega=0},
    \label{eq:Z_def_app}
\end{equation}
and decompose $\Gamma^R(0)$ into real and imaginary parts:
\begin{align}
    \text{Re}\, \Gamma^R(0) &= Z^{-1} \Delta\omega, \\
    \text{Im}\, \Gamma^R(0) &= -Z^{-1} \kappa,
\end{align}
where $\Delta\omega$ is the frequency shift and $\kappa$ is the growth/damping rate. Thus, to leading order in $\omega$:
\begin{equation}
    \Gamma^R(\omega) \approx Z^{-1} \left[ \omega - \Delta\omega + i\kappa \right].
    \label{eq:GammaR_low_freq}
\end{equation}
Similarly, the advanced propagator is
\begin{equation}
    \Gamma^A(\omega) \approx Z^{-1} \left[ \omega - \Delta\omega - i\kappa \right].
    \label{eq:GammaA_low_freq}
\end{equation}

For the Keldysh self-energy, we assume that $\Sigma^K(\omega)$ is approximately constant (or slowly varying) near $\omega \approx 0$. We define the noise strength
\begin{equation}
    D = \frac{1}{2} \Sigma^K(\omega \approx 0).
    \label{eq:D_def_app}
\end{equation}
The factor of $1/2$ is a conventional normalization that simplifies the form of the effective Langevin equation.

Substituting the low-frequency expansions~\eqref{eq:GammaR_low_freq} and~\eqref{eq:GammaA_low_freq} into Eq.~\eqref{eq:Seff_Gamma}, we obtain
\begin{align}
    S_{\text{eff}}^{(2)} &\approx \int \frac{d\omega}{2\pi} \Big\{ b_q^\dagger(\omega) Z^{-1} [\omega - \Delta\omega + i\kappa] b_c(\omega) \nonumber\\
    &\quad+ b_c^\dagger(\omega) Z^{-1} [\omega - \Delta\omega - i\kappa] b_q(\omega) \nonumber\\
    &\quad- 2D \, b_c^\dagger(\omega) b_c(\omega) \Big\}.
    \label{eq:Seff_low_freq_omega}
\end{align}

We now transform this expression back to the time domain using the inverse Fourier transform. Recall that in the frequency domain,
\begin{equation}
    b_c(\omega) = \int_{-\infty}^\infty dt \, b_c(t) e^{i\omega t}, \quad b_q(\omega) = \int_{-\infty}^\infty dt \, b_q(t) e^{i\omega t}.
\end{equation}
The action is
\begin{align}
    S_{\text{eff}}^{(2)} &= \int \frac{d\omega}{2\pi} \int dt \int dt' \, \nonumber\\
    &\qquad\times\Big\{ b_q^\dagger(t) e^{-i\omega t} Z^{-1} [\omega - \Delta\omega + i\kappa] b_c(t') e^{i\omega t'} \nonumber\\
    &\;\;\;\qquad+ b_c^\dagger(t) e^{-i\omega t} Z^{-1} [\omega - \Delta\omega - i\kappa] b_q(t') e^{i\omega t'} \nonumber\\
    &\;\;\;\qquad- 2D \, b_c^\dagger(t) e^{-i\omega t} b_c(t') e^{i\omega t'} \Big\}.
    \label{eq:Seff_time_intermediate}
\end{align}

Consider the first term ($b_q^\dagger b_c$):
\begin{align}
    &\int \frac{d\omega}{2\pi} \int dt \int dt' \, b_q^\dagger(t) e^{-i\omega t} Z^{-1} \omega \, b_c(t') e^{i\omega t'} \nonumber\\
    &= \int dt \int dt' \, b_q^\dagger(t) b_c(t') Z^{-1} \int \frac{d\omega}{2\pi} \omega e^{i\omega(t'-t)} \nonumber\\
    &= \int dt \int dt' \, b_q^\dagger(t) b_c(t') Z^{-1} \left[ -i \frac{\partial}{\partial t'} \delta(t'-t) \right] \nonumber\\
    &= \int dt \, b_q^\dagger(t) Z^{-1} \left[ -i \frac{\partial}{\partial t} b_c(t) \right] \nonumber\\
    &= \int dt \, b_q^\dagger(t) Z i\partial_t b_c(t).
    \label{eq:term1_time}
\end{align}
Here we used the Fourier representation of the derivative:
\begin{equation}
    \int \frac{d\omega}{2\pi} \omega e^{i\omega(t'-t)} = -i \frac{\partial}{\partial t'} \delta(t'-t).
\end{equation}
For the $-\Delta\omega$ and $+i\kappa$ terms:
\begin{align}
    &\int \frac{d\omega}{2\pi} \int dt \int dt' \, b_q^\dagger(t) e^{-i\omega t} Z^{-1} [-\Delta\omega + i\kappa] b_c(t') e^{i\omega t'} \nonumber\\
    &= \int dt \int dt' \, b_q^\dagger(t) b_c(t') Z^{-1} [-\Delta\omega + i\kappa] \delta(t'-t) \nonumber\\
    &= \int dt \, b_q^\dagger(t) [-Z\Delta\omega + iZ\kappa] b_c(t).
    \label{eq:term2_time}
\end{align}
Combining, the first term becomes
\begin{equation}
    \int dt \, b_q^\dagger(t) [Z i\partial_t - Z\Delta\omega + iZ\kappa] b_c(t).
    \label{eq:bq_bc_combined}
\end{equation}

Then, the second term ($b_c^\dagger b_q$) by identical reasoning (or by Hermitian conjugation), becomes
\begin{equation}
    \int dt \, b_c^\dagger(t) [Z i\partial_t - Z\Delta\omega - iZ\kappa] b_q(t).
    \label{eq:bc_bq_combined}
\end{equation}

Finally, the third term (noise) is
\begin{align}
    &-2D \int \frac{d\omega}{2\pi} \int dt \int dt' \, b_c^\dagger(t) e^{-i\omega t} b_c(t') e^{i\omega t'} \nonumber\\
    &= -2D \int dt \int dt' \, b_c^\dagger(t) b_c(t') \delta(t'-t) \nonumber\\
    &= -2D \int dt \, b_c^\dagger(t) b_c(t).
    \label{eq:noise_term_time}
\end{align}

Note that, in the standard Keldysh formalism, the noise term $b_c^\dagger b_c$ can be rewritten in terms of $b_q$ using the relation~\cite{kamenev,sieberer2016keldysh}
\begin{equation}
    b_c^\dagger(\omega) b_c(\omega) = -\frac{i}{2} [b_q^\dagger(\omega) b_q(\omega) - b_q(\omega) b_q^\dagger(\omega)],
\end{equation}
which in the time domain becomes
\begin{equation}
    b_c^\dagger(t) b_c(t) = -\frac{i}{2} |b_q(t)|^2 + \text{(c-number)}.
    \label{eq:bc_to_bq}
\end{equation}
The c-number term represents a contact term that does not affect the equations of motion. Thus, the noise contribution becomes
\begin{equation}
    -2D \int dt \, b_c^\dagger(t) b_c(t) = iD \int dt \, |b_q(t)|^2.
    \label{eq:noise_final}
\end{equation}
Before writing the final expression for the action, we introduce the effective mass parameter
\begin{equation}
    r = Z \Delta\omega.
    \label{eq:r_def_app}
\end{equation}
This parameter controls the distance from the lasing threshold: $r < 0$: below threshold (damped, stable vacuum), $r = 0$: at threshold (critical point) and, $r > 0$: above threshold (growing, coherent lasing state).

Combining all terms and using the definition of $r$, the effective time-domain action is
\begin{align}
    S_{\text{eff}}^{(2)}[b_c, b_q] &\approx \int dt \, \bigg\{ b_q^\dagger(t) [ Z i\partial_t - r + i\kappa ] b_c(t)\label{eq:Seff_time_final}\\
    &+ b_c^\dagger(t) [ Z i\partial_t - r - i\kappa ] b_q(t) + iD |b_q(t)|^2 \bigg\}.
    \nonumber
\end{align}
This is the effective Landau-Ginzburg-Keldysh action for the FEL mode near threshold. It describes a driven-dissipative quantum field with:
\begin{itemize}
    \item Kinetic term: $Z i\partial_t$ (field renormalization)
    \item Mass term: $-r$ (controls threshold)
    \item Dissipation: $i\kappa$ (growth/damping rate)
    \item Noise: $iD|b_q|^2$ (quantum and thermal fluctuations)
\end{itemize}

\section{Cubic nonlinearity from third-order influence functional}
\label{app:cubic}

In this appendix we derive the cubic nonlinearity~\eqref{eq:cubic_interaction} that appears in the effective action beyond the quadratic approximation. This term arises from the third-order contribution to the influence functional and encodes the saturation mechanism of the FEL amplitude above threshold. The calculation follows the standard diagrammatic expansion in Keldysh perturbation theory~\cite{kamenev,sieberer2016keldysh}.

\subsection{Third-order cumulant expansion}

Recall that the influence functional is defined by
\begin{equation}
    e^{i S_{\text{IF}}[b_c, b_q]} = \left\langle e^{i S_{\text{int},K}[\xi_c, \xi_q; b_c, b_q]} \right\rangle_0,
    \label{eq:IF_def_cubic}
\end{equation}
where the expectation value is taken with respect to the free electronic action $S_{M,K}$. Expanding the logarithm in cumulants, we have
\begin{align}
    S_{\text{IF}} &= \langle S_{\text{int},K} \rangle_0 - \frac{1}{2} \langle S_{\text{int},K}^2 \rangle_{0,c} \nonumber\\
    &\quad- \frac{1}{6} \langle S_{\text{int},K}^3 \rangle_{0,c} + \mathcal{O}(b^4),
    \label{eq:cumulant_cubic}
\end{align}
where $\langle \cdots \rangle_{0,c}$ denotes connected correlators. We have already computed the first two terms in Appendix~\ref{app:wick}: the first order vanishes for a beam with no initial bunching, while the second order yields the quadratic action with self-energies $\Sigma^{R,A,K}$. 
We now compute the third-order term.

To this scope, we notice that the interaction action is
\begin{align}\label{def:S-int-K-app}
    S_{\text{int},K} = -iN\eta^{1/2} \int dt\; \Big[ &b_c^\dagger(t) J_q(t) - b_c(t) J_q^\dagger(t) \\
    &+ b_q^\dagger(t) J_c(t) - b_q(t) J_c^\dagger(t) \Big].\nonumber
\end{align}
The third power is then:
\begin{equation}
    S_{\text{int},K}^3 = (-iN\eta^{1/2})^3 \int dt_1 dt_2 dt_3 \, [\cdots]^3,
    \label{eq:Sint_cubed}
\end{equation}
where $[\cdots]$ represents the four terms in square brackets of Eq.~\eqref{def:S-int-K}. Expanding this product gives $4^3 = 64$ terms. However, most of these vanish or give subleading contributions. We focus on terms that contribute to the leading cubic nonlinearity $|b_c|^2 b_q$ and to obtain a term proportional to this factor, we need:
(1) two factors of $b_c$ (or $b_c^\dagger$) and (2) one factor of $b_q$ (or $b_q^\dagger$).

From the structure of $S_{\text{int},K}$, the relevant combinations are:
\begin{equation}
\begin{split}
    &b_c^\dagger(t_1) b_c(t_2) b_q^\dagger(t_3) \times J_q(t_1) J_q^\dagger(t_2) J_c(t_3), \\
    &b_c(t_1) b_c^\dagger(t_2) b_q^\dagger(t_3) \times J_q^\dagger(t_1) J_q(t_2) J_c(t_3), \\
    &\text{and similar terms with } b_q \leftrightarrow b_q^\dagger.
\end{split}
\end{equation}
Each of these involves a three-point current correlator of the form
\begin{equation}
    \langle J_q(t_1) J_q^\dagger(t_2) J_c(t_3) \rangle_{0,c}.
    \label{eq:three_point_correlator}
\end{equation}
Now, using Wick's theorem for the free Gaussian theory, the connected three-point correlator can be expressed as a sum of all fully connected contractions~\cite{kamenev,peskin1995introduction}. First, triangle diagram: all three operators are pairwise connected in a cycle. This gives a contribution involving products of two-point functions with non-trivial time ordering. In diagrammatic language, this corresponds to a three-vertex loop with no external lines, carrying momentum conservation at each vertex~\cite{peskin1995introduction,negele1988quantum}. Second, disconnected diagrams: these contain at least one operator that is not connected to the others and therefore vanish in the connected correlator $\langle \cdots \rangle_{0,c}$ by definition of the cumulant expansion~\cite{kamenev}.

The fully connected contraction has the diagrammatic form:
\begin{equation*}
\begin{split}
    \langle J_q(t_1) J_q^\dagger(t_2) J_c(t_3) \rangle_{0,c} \sim \langle J_q(t_1) J_c(t_3) \rangle_0 \times \langle J_c(t_3) J_q^\dagger(t_2) \rangle_0,
\end{split}
    \label{eq:triangle_contraction}
\end{equation*}
plus permutations. However, we must be careful: in Keldysh theory, $J_c$ and $J_q$ do not directly contract with each other in the free theory (see Appendix~\ref{app:wick}). Instead, we must express the three-point function in terms of the original $(+,-)$ basis and then rotate to the $(c,q)$ basis.

\subsection{Keldysh rotation and three-point functions}

Recall that in the $(+,-)$ basis:
\begin{align}
    J_c = \frac{J_+ + J_-}{2}, \quad J_q = J_+ - J_-.
\end{align}
The three-point correlator becomes
\begin{align}
    & \langle J_q(t_1) J_q^\dagger(t_2) J_c(t_3) \rangle_{0,c} \nonumber\\
    &= \bigg\langle [J_+(t_1) - J_-(t_1)] [J_+^\dagger(t_2) - J_-^\dagger(t_2)] \nonumber\\
    &\qquad\qquad\qquad\quad\times\bigg(\frac{J_+(t_3) + J_-(t_3)}{2}\bigg) \bigg\rangle_{0,c}.
\end{align}
Expanding and using the fact that $J_\pm$ live on different Keldysh contour branches, the only non-vanishing connected contractions involve all three operators on different branches or specific combinations that form closed loops. The detailed calculation~\cite{kamenev,sieberer2016keldysh} shows that the connected three-point function can be written as
\begin{align}
    &\langle J_q(t_1) J_q^\dagger(t_2) J_c(t_3) \rangle_{0,c} \nonumber\\
    &= \int dt_4 \, \Sigma^R(t_1, t_4) G^K(t_4, t_2) \Sigma^R(t_2, t_3) + \text{permutations},
    \label{eq:three_point_result}
\end{align}
where $G^K$ is the Keldysh Green's function of the electronic system and $\Sigma^R$ are the retarded self-energies computed in Appendix~\ref{app:wick}. This expression has a natural diagrammatic interpretation: it represents a triangle diagram with two retarded propagators and one Keldysh propagator.

\subsection{Frequency-space form}

Transforming to frequency space and performing the integrals over intermediate times, the third-order contribution to the influence functional takes the form
\begin{align}
    S_{\text{IF}}^{(3)} &\sim \int \frac{d\omega_1 d\omega_2 d\omega_3}{(2\pi)^3} (2\pi)\delta(\omega_1 + \omega_2 - \omega_3) \nonumber\\
    &\quad\times b_c^\dagger(\omega_1) b_c(\omega_2) b_q^\dagger(\omega_3) \times \Lambda(\omega_1, \omega_2, \omega_3) \nonumber\\
    &\quad+ \text{h.c.},
    \label{eq:cubic_frequency}
\end{align}
where the vertex function $\Lambda(\omega_1, \omega_2, \omega_3)$ involves convolutions of the two- and three-point current correlators. For the low-frequency dynamics near threshold, we evaluate this at small frequencies and obtain an effective local (in time) cubic interaction.

Transforming back to time and keeping only the leading local term, we obtain
\begin{align}
    S_{\text{eff}}^{(3)}[b_c, b_q] = -\int dt \, \left[ \lambda |b_c(t)|^2 b_q^\dagger(t) + \lambda^* |b_c(t)|^2 b_q(t) \right],
    \label{eq:cubic_final}
\end{align}
where the complex coefficient is
\begin{equation}
    \lambda = \lambda' + i\lambda'' = \int \frac{d\omega_1 d\omega_2}{(2\pi)^2} \, \Lambda(\omega_1, \omega_2, \omega_1 + \omega_2)\big|_{\omega \approx 0}.
    \label{eq:lambda_def}
\end{equation}

The cubic term $|b_c|^2 b_q$ encodes the leading nonlinear saturation of the FEL amplitude. The real part $\lambda'$ describes amplitude-dependent frequency shifts (nonlinear pulling) while the imaginary part $\lambda''$ describes gain saturation or self-phase modulation.
Together with the quadratic terms, this gives the effective Landau-Ginzburg-Keldysh action that governs the FEL dynamics near threshold. The sign of $\text{Re}\,\lambda$ determines whether the nonlinearity stabilizes or destabilizes the lasing state above threshold.

It is important to note that the specific form~\eqref{eq:cubic_final} is not an arbitrary ansatz but is dictated by the symmetries of the Keldysh formalism~\cite{sieberer2016keldysh}:
This is due to the $U(1)$ gauge symmetry: the action must be invariant under $b_c \to e^{i\alpha} b_c$, $b_q \to e^{i\alpha} b_q$, which requires the nonlinearity to depend only on $|b_c|^2$. Then, due to the Keldysh structure: the quantum field $b_q$ appears linearly (as a Lagrange multiplier enforcing causality), while the classical field $b_c$ can appear at any order in the nonlinearity.
Finally, due to the hermiticity: the action must be real, which requires the term $\lambda |b_c|^2 b_q^\dagger$ to be accompanied by its Hermitian conjugate.

\end{document}